\documentclass{aa}   
\usepackage{xcolor}
\usepackage{graphicx}
\usepackage{txfonts}
\usepackage{lipsum}
\usepackage{subcaption}         
\usepackage{lscape}             
\usepackage{placeins}           
\usepackage{natbib,twoopt}

\usepackage[breaklinks=true]{hyperref}
\bibpunct{(}{)}{;}{a}{}{,}
\makeatletter
\newcommandtwoopt{\citeads}[3][][]{\href{http://adsabs.harvard.edu/abs/#3}%
{\def\hyper@linkstart##1##2{}%
\let\hyper@linkend\@empty\citealp[#1][#2]{#3}}}
\newcommandtwoopt{\citepads}[3][][]{\href{http://adsabs.harvard.edu/abs/#3}%
{\def\hyper@linkstart##1##2{}%
\let\hyper@linkend\@empty\citep[#1][#2]{#3}}}
\newcommandtwoopt{\citetads}[3][][]{\href{http://adsabs.harvard.edu/abs/#3}%
{\def\hyper@linkstart##1##2{}%
\let\hyper@linkend\@empty\citet[#1][#2]{#3}}}
\newcommandtwoopt{\citeyearads}[3][][]%
{\href{http://adsabs.harvard.edu/abs/#3}
{\def\hyper@linkstart##1##2{}%
\let\hyper@linkend\@empty\citeyear[#1][#2]{#3}}}
\makeatother

\hypersetup{
  colorlinks=true,
  urlcolor=blue,
  linkcolor=blue,
  citecolor=blue
}

\begin{document}
   \title{Astrometric properties of reference frame sources as a function of redshift}
   \titlerunning{Astrometric properties of reference frame sources as a function of redshift}      
   \authorrunning{Zhang et al.}    

   \author{Zhiyun Zhang\inst{1,2}
        \and N. Liu\inst{1,2}\fnmsep\thanks{Corresponding author; niu.liu@nju.edu.cn}
        \and Xiaxuan Zhang\inst{1,2}
        \and I. Nurul Huda\inst{1,2,3}
        \and Sufen Guo\inst{4}
        \and Z. Zhu\inst{1,2,5}
        \and J.-C. Liu\inst{1,2}
        \and J. Yao\inst{1,2}
        \and Z.-W. Wang\inst{1,2}
        \and H.-F. Yu\inst{1,2}
        \and D.-D. Zhang\inst{1,2}
        }

   \institute{School of Astronomy and Space Science, Nanjing University, Nanjing 210023, China
   \and Key Laboratory of Modern Astronomy and Astrophysics (Ministry of Education), Nanjing University, Nanjing 210023, China
   \and Research Center for Computing, National Research and Innovation Agency, Bandung 40135, Indonesia
   \and School of Physical Science and Technology, Xinjiang University, Urumqi, 830046, China
   \and University of Chinese Academy of Sciences, Nanjing, 211135, China}

   \date{Received September 30, 20XX}
 
  \abstract
   {Previous studies based on the latest realisation of the International Celestial Reference Frame (ICRF3) have suggested a correlation between astrometric properties (such as the radio-optical offset) and redshift for active galactic nuclei (AGNs).}
   {We extend these investigations by using a large, all-sky sample of approximately 22\,000 compact radio sources from the Radio Fundamental Catalogue (RFC) to examine this relationship in a systematic and statistically robust manner.} 
   {We compiled redshifts for about 10\,000 RFC sources over the range $0\!<\!z\!<\!5$ by combining data from the Dark Energy Spectroscopic Instrument Data Release 1 and the Sloan Digital Sky Survey Data Release 17/19 with additional datasets from the NASA/IPAC Extragalactic Database. 
    Cross-matching with \textit{Gaia} Data Release 3 yielded a sample of 4\,068 RFC objects with reliable spectroscopic redshifts and classifications, including galaxies and quasi-stellar objects (QSOs).
    We analysed the redshift dependence of their radio astrometric properties from very long baseline interferometry (VLBI) and their optical astrometric properties from \textit{Gaia}.}
   {We find that the VLBI astrometric properties show no significant dependence on redshift within the achieved level of precision.
   In contrast, several optical astrometric quantities exhibit clear redshift-dependent behaviour.
   The median absolute radio–optical offsets decrease markedly over $0\!<z\!<0.5$, where galaxies dominate the sample, decline more gradually over $0.5\!<z\!<1.3$, and exhibit a mild increase at $z\!>\!1.3$, where QSOs dominate.
   Similar behaviour is observed for several \textit{Gaia} astrometric quantities, including astrometric uncertainties, proper motions, and $G$ magnitudes. 
   These behaviours can be largely explained by the dependence of \textit{Gaia} astrometric performance on $G$ magnitude and by the evolution of the $G$ magnitude with redshift.
   }
   {}

   \keywords{astrometry -- 
   galaxies: distances and redshifts -- 
   catalogs -- 
   reference systems.}
   
   \maketitle

\section{Introduction}
Accurate celestial reference frames are fundamental to modern astrometry. 
These frames are constructed through observations of extragalactic active galactic nuclei (AGNs) with compact radio cores, which provide the most stable reference points.
A prominent example is the International Celestial Reference Frame (ICRF), which is established using geodetic very long baseline interferometry (VLBI) to measure the radio positions of thousands of compact AGNs with sub-milliarcsecond (mas) precision.
The latest version of the ICRF \citepads[ICRF3;][]{2020A&A...644A.159C} includes 4\,536 objects at $S/X$ bands (2.3/8.4~GHz), 824 objects at $K$ band (24~GHz) and 678 objects at $X/Ka$ bands (8.4/32~GHz).
With the advent of the \textit{Gaia} mission \citepads{2016A&A...595A...1G}, particularly the release of \textit{Gaia}~DR3 \citepads{2023A&A...674A...1G}, an optical celestial reference frame with precision comparable to that of VLBI has become available (\citeads{2021A&A...649A...2L}; \citeads{2022A&A...667A.148G}).
For the majority of ICRF3 sources, the \textit{Gaia} and VLBI positions are consistent within the expected uncertainties, typically at the mas level.
However, at least $6\%$ of sources with both VLBI and \textit{Gaia} positions show statistically significant offsets at the $99\%$ confidence level (\citeads{2016A&A...595A...5M}; \citeads{2017MNRAS.467L..71P}).

A number of recent studies have investigated the origin and behaviour of these radio–optical offsets.
They have shown that radio–optical offsets correlate with various physical parameters, 
such as AGN variability (\citeads{2022ApJ...939L..32S}; \citeads{2024A&A...684A..93L}), colour index \citepads{2025A&A...695A.135L}, radio structure \citepads{2025AJ....169..173X}, and jet orientation (\citeads{2017A&A...598L...1K}, \citeyearads{2020MNRAS.493L..54K}; \citeads{2019ApJ...871..143P}; \citeads{2019MNRAS.482.3023P}; \citeads{2021A&A...651A..64L}). 
Some works have also reported a dependence on redshift (\citeads{2017ApJ...835L..30M},  \citeyearads{2019ApJ...873..132M}; \citeads{2021A&A...652A..87L}), 
suggesting possible evolutionary or observational effects.
Moreover, several \textit{Gaia} astrometric uncertainties, including those of parallax, proper motion and source position, depend strongly on the $G$ magnitude \citepads{2023A&A...674A...1G}. 
Since the $G$ magnitude varies systematically with redshift \citepads{2025A&A...695A.135L}, these uncertainties may also show a corresponding dependence on redshift.
Additionally, redshift provides a direct measure of cosmic time and is a fundamental parameter for studying the physical nature and evolution of AGNs \citepads{2024AJ....167...62D} and kinematic structure of the Universe \citepads{2025NatAs...9.1396M}.
Thus, studying the astrometric properties of reference sources as a function of redshift contributes to the evaluation of systematic effects and the improvement of the accuracy and long-term stability of the celestial reference frame.

Recent studies have explored the connection between the physical properties, redshifts, and astrometric behaviour of reference sources (\citeads{2019ApJ...873..132M}; \citeads{2021A&A...652A..87L}; \citeads{2022ApJ...939L..32S}; \citeads{2023A&A...674A...1G}; \citeads{2024ApJS..274...27M}; \citeads{2024A&A...684A..93L}; \citeads{2025A&A...695A.135L}; \citeads{2025MNRAS.542.2389H}).
In particular, \citetads{2022ApJS..260...33S} presented a detailed spectroscopic characterisation of a subset of ICRF3 sources based on Sloan Digital Sky Survey Data Release 16 \citepads[SDSS~DR16;][]{2020ApJS..249....3A}.
This work indicates the importance of reliable redshifts and spectral classifications for understanding the physical nature of reference frame objects and their astrometric behaviour.
Extending such analyses to larger samples can improve our understanding of the redshift dependence of astrometric properties.

With more than four times the number of sources in ICRF3, the Radio Fundamental Catalogue \citepads[RFC;][]{2025ApJS..276...38P} provides a large all-sky sample of compact radio sources, enabling robust statistical studies of astrometric properties as a function of redshift.
At present, while the RFC cross-match table includes redshift and classification information for a substantial fraction of sources, this information remains incomplete and non-uniform.
Only $\sim\!45\%$ of the RFC sources have literature-based data from the NASA/IPAC Extragalactic Database \citepads[NED;][]{1991ASSL..171...89H}.
These literature-based redshifts differ in precision because they are obtained from different methods, such as spectroscopy (\citeads{2009MNRAS.399..683J}; \citeads{2017ApJS..233...25A}), photometry (\citeads{2009ApJS..180...67R}; \citeads{2014ApJS..210....9B}), and machine-learning estimates (\citeads{2022ApJS..259...55N}; \citeads{2023ApJS..264....4M}).
Similarly, literature-based classifications rely on non-uniform criteria across different studies.

A comparable situation applies to the Optical Characteristics of Astrometric Radio Sources catalogue (OCARS; \citeads{2016ARep...60..996M}; \citeyearads{2018ApJS..239...20M}; \citeyearads{2025AJ....170...48M}) and the Large Quasar Astrometric Catalogue (LQAC; \citeads{2009A&A...494..799S}, \citeyearads{2012A&A...537A..99S}, \citeyearads{2015A&A...583A..75S}; \citeads{2018A&A...614A.140G}; \citeads{2019A&A...624A.145S}, \citeyearads{2024A&A...683A.112S}), which includes most RFC sources and whose redshifts are likewise primarily compiled from NED, the SIMBAD astronomical database \citepads{2000A&AS..143....9W}, and SDSS.
The rapid development of large spectroscopic surveys now enables a substantial improvement.
The latest data releases from the Dark Energy Spectroscopic Instrument \citepads[DESI~DR1;][]{2025arXiv250314745D} and the SDSS \citepads[SDSS~DR19;][]{2025arXiv250707093S} provide high-quality spectra and precise spectroscopic redshifts for a much larger fraction of RFC objects.
Based on these spectroscopic data, the present work provides an independent compilation of spectroscopic redshifts based directly on SDSS and DESI data and supplements the RFC cross-match table with updated and more homogeneous redshift and classification information for RFC sources.

In this work, we provide the largest RFC-based spectroscopic-redshift sample cross-matched with \textit{Gaia}~DR3, based on the latest spectroscopic data from DESI~DR1 and SDSS~DR17/DR19, and systematically analyse redshift trends of both radio and optical astrometric properties, separating galaxies and quasi-stellar objects (QSOs).
The paper is organised as follows.
In Sect.~\ref{sec:Data and Sample Selection}, we describe the RFC catalogue, DESI/SDSS spectroscopic data, and our sample selection procedures. 
In Sect.~\ref{sec:Result}, we analyse the redshift dependence of astrometric parameters for the total sample and for different source classes.
We discuss the physical origins of the redshift dependence of the astrometric parameters and compare our results with previous studies in Sect.~\ref{sec:Discussion}.
Finally, we present our conclusions in Sect.~\ref{sec:Conclusion}.

\section{Data and sample selection}    \label{sec:Data and Sample Selection}

\subsection{The Radio Fundamental Catalogue}

In this work, we adopted the release version rfc\_2025c of the RFC catalogue\footnote{\url{https://astrogeo.org/sol/rfc/rfc_2025c/}} as the parent sample for redshift determination and cross-matching with \textit{Gaia}~DR3 data.
This release contains 21\,949 sources, whose absolute positions and correlated flux densities were determined from the analysis of virtually all publicly available VLBI sessions conducted over the past four decades.
The catalogue achieves milliarcsecond-level astrometric accuracy by applying a novel technique that combines and calibrates heterogeneous VLBI datasets, and by rigorously evaluating systematic and random errors through decimation tests, cross-comparisons, and error modelling \citepads{2025ApJS..276...38P}.
Its high precision and wide sky coverage make it fundamental to a wide range of applications, from celestial reference frame definition to space geodesy, spacecraft navigation, AGN studies, and modern VLBI phase-referencing observations.

\subsection{DESI/SDSS spectroscopic data}
In this work, we primarily used spectroscopic data from the first data release of DESI \citepads[DESI~DR1;][]{2019AJ....157..168D,2025arXiv250314745D}, a next-generation wide-field survey installed on the 4-metre Mayall Telescope at Kitt Peak National Observatory, to determine redshifts.
DESI is equipped with 5\,000 robotic fibre positioners and a 3.2° field of view, enabling highly efficient simultaneous observations of thousands of targets. 
Its spectrographs provide continuous wavelength coverage from $3\,600~\AA$ to $9\,800~\AA$ with a typical resolving power of $R\!\sim\!2\,000$–$5\,000$, allowing precise measurements of spectral features across a broad range of astrophysical sources. 
The main survey of DESI~DR1 provides high-confidence redshifts for approximately 18.7 million objects, including 13.1 million galaxies, 1.6 million quasars, and 4 million stars, making it the largest extragalactic redshift sample ever assembled.
We show DESI spectra and their high-precision redshifts for three example sources in Appendix~\ref{appendix:desi-spectra}: a galaxy exhibiting Balmer emission lines ($z\!<\!0.6$), a QSO showing Mg\,II, C\,III] and C\,IV lines ($z\!<\!2.1$), and a QSO showing Ly$\alpha$ emission line with Ly$\alpha$ forest ($z\!>\!2.1$).
In all three cases, the redshift uncertainties are at the level of $\sim\!10^{-4}$.

We also made use of spectroscopic data from the seventeenth data release of SDSS \citepads[SDSS~DR17;][]{2022ApJS..259...35A}, the final release of the SDSS-IV programme.
SDSS~DR17 provides one of the most comprehensive spectroscopic datasets available, obtained with the 2.5~m Sloan Foundation Telescope at Apache Point Observatory. 
Compared to earlier releases such as DR13, SDSS~DR17 represents a major expansion and refinement of the spectroscopic dataset. 
The number of reliable spectra in the main QSO sample from the Extended Baryon Oscillation Spectroscopic Survey \citepads[eBOSS;][]{2016AJ....151...44D} has increased from roughly $34\,000$ in DR13 to more than $430\,000$ in DR17. 
Similarly, the spectroscopic sample of LRGs and ELGs has expanded from about $47\,000$ to over $460\,000$ objects.
DR17 also incorporates significant methodological improvements, including the use of PCA-based spectral fitting and extensive visual inspection in addition to the standard SDSS pipeline, resulting in more accurate classifications, particularly for quasars. 
As the final data release of SDSS-IV, DR17 thus provides a more complete and higher-quality spectroscopic dataset than previous releases, enabling more robust studies of galaxies, quasars, and the large-scale structure of the Universe.

In addition, we also made use of spectroscopic data from the latest SDSS~DR19 \citepads{2025arXiv250707093S}, which is part of the SDSS-V program.
As the first substantial public release of new data from the SDSS-V ecosystem, DR19 provides updates relative to the final SDSS-IV release. 
The DR19 data include early observations from the Black Hole Mapper (BHM), which provide expanded spectroscopic coverage of quasars and AGN.
The new data and improved measurements of DR19 complement the DR17 dataset, allowing for more robust analyses.

\begin{figure*}
  \centering
  \includegraphics[width=0.9\textwidth]{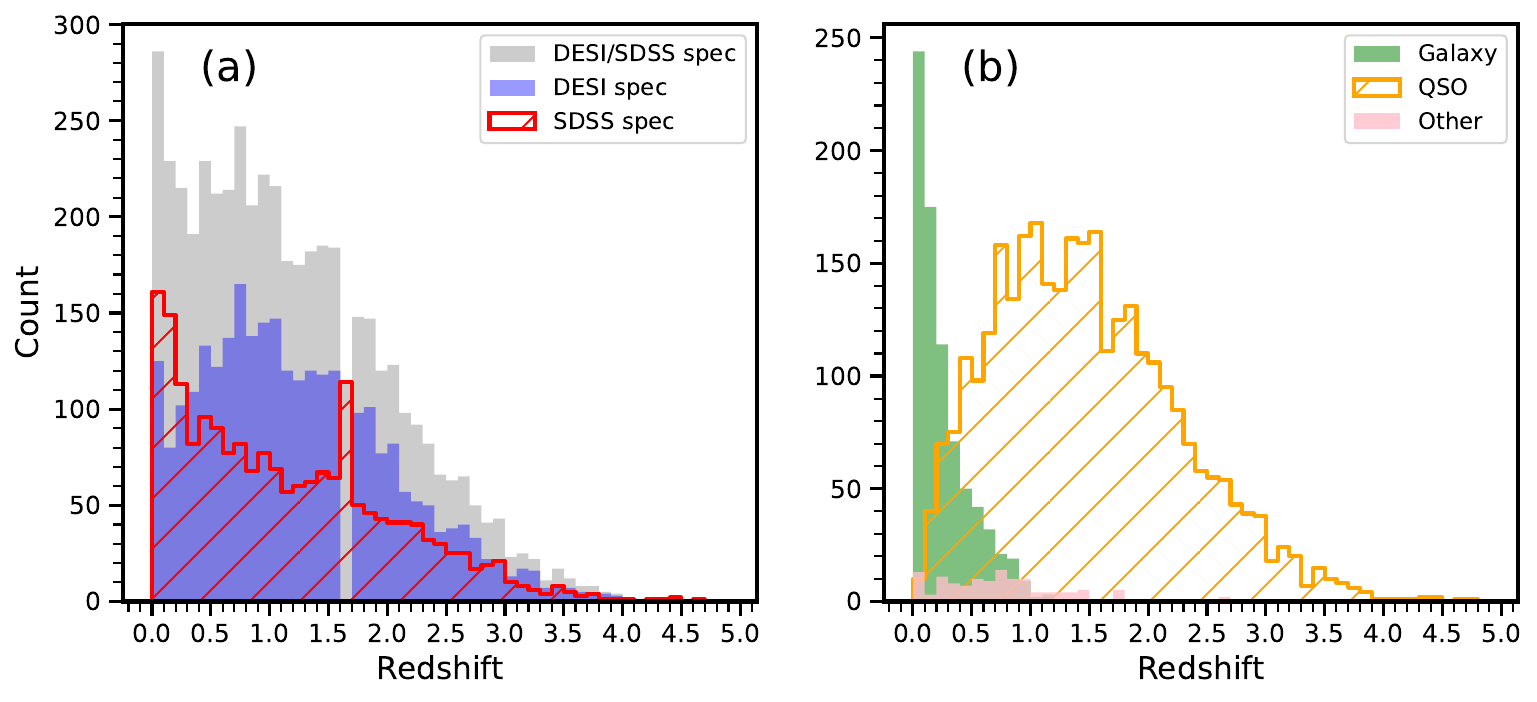}  
  \caption[]{\label{fig:zcount} 
  Redshift distribution of objects in this work. 
  Panel a: The grey histogram shows sources with spectroscopic redshifts from DESI/SDSS, the blue and red histograms show the distributions of DESI and SDSS redshifts, respectively.
  Panel b: The same sample as panel a separated by spectral classification: green for galaxies, orange for QSOs, and pink for sources classified as stars or lacking a reliable classification.
  }
\end{figure*}

\subsection{Sample selection}
The redshift information for RFC sources was assembled in a hierarchical manner during the sample selection process. 
We adopted a sequential strategy that prioritised the most reliable spectroscopic measurements. 
First, we used spectroscopic redshifts from the DESI~DR1 catalogue with quality flag ${\tt ZWARN}=0$, resulting in 3\,403 reliable redshifts. 
For sources without reliable DESI redshifts, we adopted spectroscopic redshifts from SDSS~DR17 and SDSS~DR19 with ${\tt ZWARNING}=0$, which provided an additional 1\,832 sources.
In total, this sequential procedure provided 5\,235 sources with reliable spectroscopic redshifts from DESI or SDSS.
All these spectroscopic redshifts were required to satisfy $z\!>\!0$ and have measurement uncertainties smaller than 0.01.

For sources lacking spectroscopic measurements, we supplemented the sample with 4,672 redshifts compiled from the literature.
Most of these literature-based redshifts were compiled from NED by the RFC cross-match table.

Following this procedure, we obtained redshift information for 9\,907 RFC sources in the catalogue. 
More than half of these redshifts were provided by the latest DESI and SDSS spectroscopic observations.
This combined dataset provided broad coverage over the redshift range $0\!<z\!<5$, ensuring both high reliability for the majority of the sample and extensive completeness for statistical analysis.

To assess the reliability of the adopted spectroscopic redshifts, we performed cross-checking between independent surveys (Appendix~\ref{appendix:z-crosscheck}). 
Overall, we found that the majority of sources show good agreement in redshifts across different methods ($|\Delta z|\!<\!0.1$).
However, we found that DESI spectroscopic redshifts in the range $1.6\!<z\!<1.7$ show significant deviations from redshifts measured by other methods.
According to \citetads{2025arXiv250314745D}, DESI spectra in the range $1.6\!<z\!<1.63$ 
are constrained only by the [O\,{\sc ii}] $\lambda\lambda3726,3729$ doublet and are affected by significant sky background, 
whereas spectra in the range $1.63\!<z\!<1.7$ have no major emission line coverage.
As a result, the DESI spectroscopic redshift solutions in this range are susceptible to unphysical fits and show systematic offsets compared with other determinations.  

To minimise the impact of such unreliable measurements, we first excluded all DESI redshifts within the range $1.6\!<z\!<1.7$ from our sample.
For the remaining sources that had both SDSS and DESI spectroscopic measurements, we retained only those for which the two redshifts were consistent within $|\Delta z|\!<\!0.1$.
The consistency criterion of $|\Delta z|\!<\!0.1$ was chosen conservatively to flag catastrophic failures rather than statistical uncertainties.
Additionally, for sources with DESI redshifts in the range $1.6\!<z\!<1.7$ but still satisfied the consistency criterion, we retained these objects and used the SDSS redshift for our analysis.
These conservative cuts ensured that our sample used in the statistical analysis was both robust and internally consistent across surveys.
After applying these selection criteria, 464 sources were excluded, leaving a sample of 4\,771 objects ($\sim\!90\%$) with reliable spectroscopic redshifts from either DESI or SDSS.
By subsequently cross-matching with \textit{Gaia} DR3, we obtained 4\,068 objects that had spectroscopic redshifts from either SDSS or DESI, including 934 ICRF3 sources, which constitute our final sample.

We are particularly interested in whether the spectral classification of RFC sources affects the relationship between their astrometric properties and spectroscopic redshift.
For this purpose, we used the spectral classifications provided by SDSS and DESI, resulting in 784 galaxies, 3\,151 QSOs, and 133 sources that were classified as stars or had inconsistent classifications between SDSS and DESI.
Based on this classification, we further examined the relationships between the astrometric parameters of RFC sources and redshift.

\subsection{Redshift distributions}

After applying the sample selection and spectral classification, we present a series of redshift distributions in Fig.~\ref{fig:zcount}. 
A complementarity is seen in the redshift distribution of RFC sources: SDSS spectroscopic redshifts are primarily concentrated at low redshifts, whereas DESI spectroscopic measurements extend the redshift coverage toward higher redshifts. 
A deficit of DESI sources appears at $1.6\!<z\!<1.7$, reflecting known reliability issues in this redshift range.
The corresponding excess of SDSS redshifts in this range arises because we used the SDSS redshift for the sources in this range whose DESI and SDSS redshifts satisfy the $|\Delta z|\!<\!0.1$ criterion in our analysis.
The distribution of the same sample separated by spectral classification shows that galaxies dominate the low-redshift population, close to the local Universe ($0\!<z\!<0.5$), whereas QSOs populate a broad redshift range extending to $z\!>\!4$, with their highest concentration occurring at $1.0\!<z\!<1.5$.

\section{Results}    \label{sec:Result}
We systematically analysed the redshift dependence of the astrometric parameters of RFC sources obtained from radio VLBI observations and optical \textit{Gaia} DR3 measurements (binned with $\Delta z\!=\!0.2$). 
All the data were smoothed using the Nadaraya–Watson estimator to better illustrate the trends of the astrometric parameters as a function of redshift. 

\begin{figure*}
  \centering
  \includegraphics[width=0.9\textwidth]{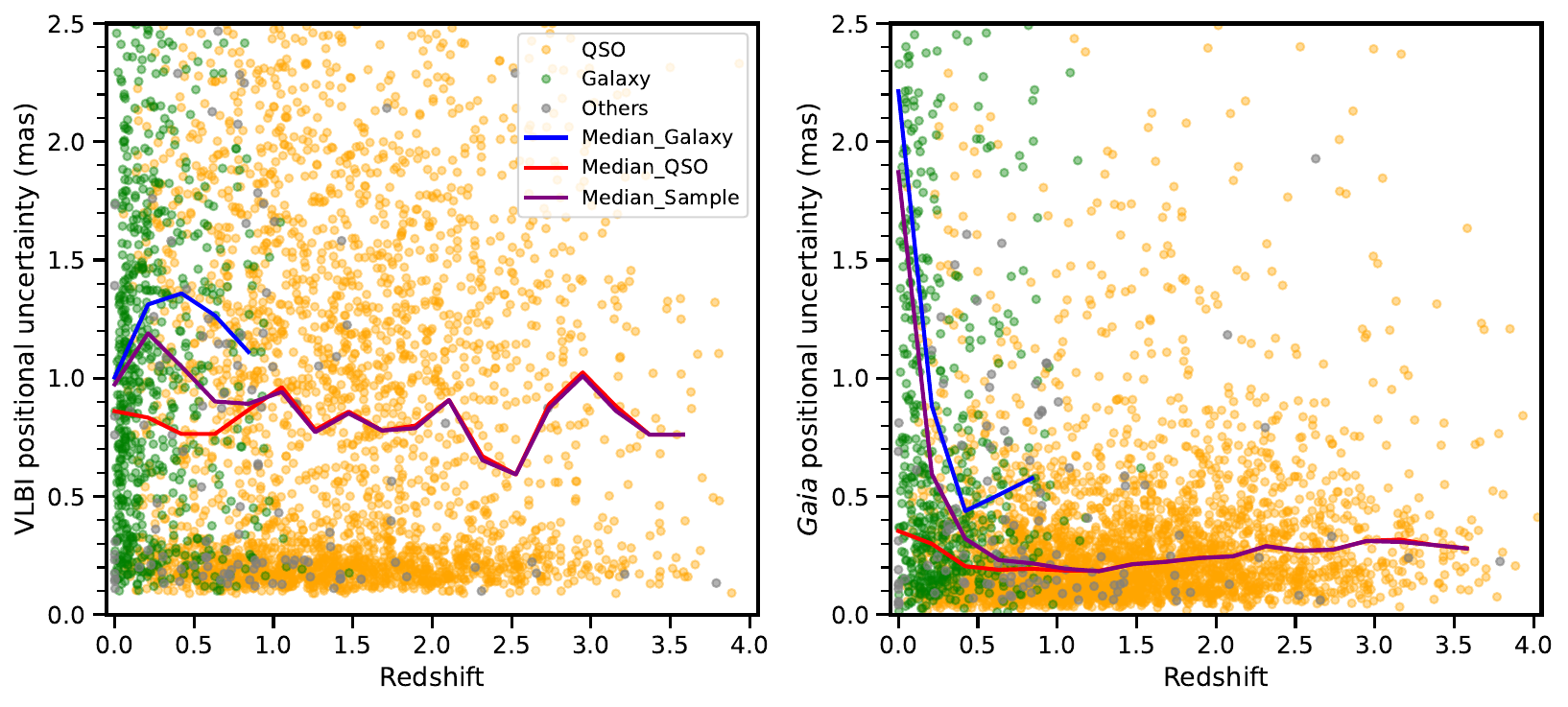}  
  \caption[]{\label{fig:uncer} 
  Positional uncertainty derived from RFC VLBI observations (left) and \textit{Gaia} optical measurements (right) as functions of redshift.
  For each panel, the orange data points indicate QSOs, the green data points indicate galaxies in the sample, and the grey data points indicate sources that are either unclassified or lack a reliable classification. 
  The blue and red solid lines represent the median values of the galaxies and QSOs, respectively, whereas the purple line represents the median value of the total sample.
  In addition, to ensure the reliability of the results, median values are not shown for bins containing fewer than 20 data points.
  }
\end{figure*}

\begin{figure*}
  \centering
  \includegraphics[width=0.9\textwidth]{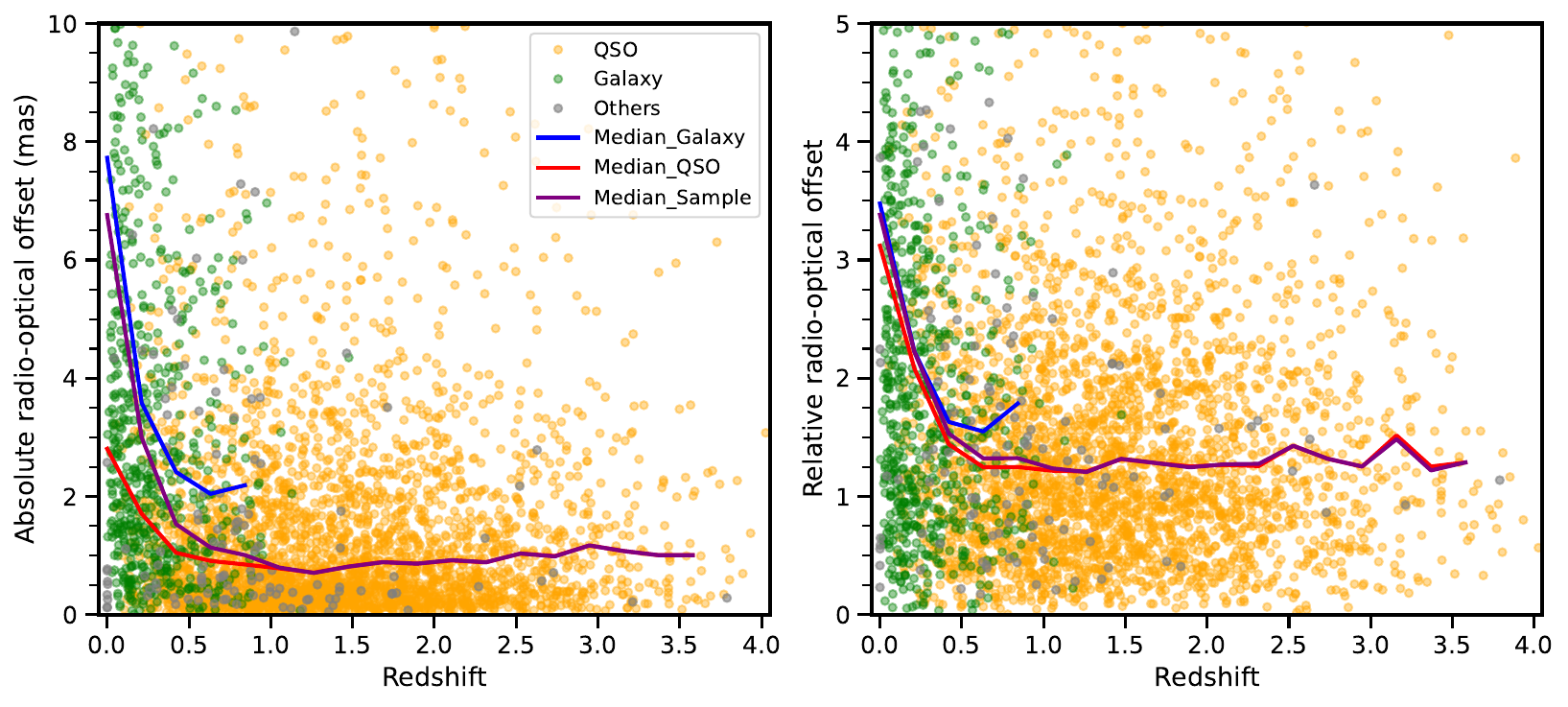}  
  \caption[]{\label{fig:offsets} 
  Absolute radio-optical offset (left) and relative radio-optical offset (right) as functions of redshift.
  The meanings of the colours, data points, and lines are the same as those in Fig.~\ref{fig:uncer}.
  }
\end{figure*}

\subsection{Astrometric positional precision}    \label{sec:Radio}
We compared the median positional uncertainties for RFC and \textit{Gaia} DR3. 
The positional uncertainties ($\sigma_{\texttt{p}}$) of VLBI and \textit{Gaia} are both defined as the semi-major axis of the positional error ellipse and are calculated as \citepads{2016A&A...595A...4L}:
\begin{equation}
  \sigma_{\texttt{p}}\!=\!\sqrt{\frac{1}{2}(\sigma_{\alpha*}^2+\sigma_{\delta}^2)+\frac{1}{2}\sqrt{(\sigma_{\alpha*}^2-\sigma_{\delta}^2)^2+(2\sigma_{\alpha*}\sigma_{\delta}\rho_{\alpha,\delta})^2}},
\end{equation}
where $\sigma_{\alpha*}\!=\!\sigma_{\alpha}\cos\delta$ and $\sigma_{\delta}$ are the formal uncertainties in right ascension and declination, respectively, and $\rho_{\alpha,\delta}$ is their correlation coefficient.
As shown in Fig.~\ref{fig:uncer}, the radio positional uncertainties of the sample show no significant variation with redshift, remaining around 0.8~mas across all redshift ranges.
In contrast, the optical positional uncertainties decrease rapidly over the range $0\!<z\!<0.5$, followed by a slower decrease over $0.5\!<z\!<1$, while remaining roughly constant at $\sim\!0.2$~mas for $z\!>\!1$.
For most RFC sources at $z\!>\!0.5$, the \textit{Gaia} optical positional uncertainties are smaller than the corresponding radio positional uncertainties.
We note that for ICRF3 sources, the median radio and optical positional uncertainties both remain roughly constant at approximately 0.2~mas across all redshifts.

We further examined the positional uncertainties of different types of RFC sources and find that QSOs at $z\!>\!0.5$ show no significant redshift dependence in either the radio or optical bands.
Galaxies have much larger positional uncertainties than QSOs in both radio and optical bands, and their optical uncertainties decrease significantly with redshift within $0\!<z\!<0.5$.
This suggests that the negative gradient in the optical positional uncertainties of the total sample within this redshift range is most likely driven by the contribution from galaxies.

We also analysed the redshift dependence of the structure index and compactness of ICRF3 sources in our sample separately at the $S$, $X$, $K$, and $Q$ bands \citepads{2000ApJS..128...17F}.
These two parameters characterise the radio morphological properties of the sources and are obtained from the Bordeaux VLBI Image Database \citepads[BVID;][]{2019evga.conf..219C}\footnote{\url{https://bvid.astrophy.u-bordeaux.fr/}}.
Since only a small number of sources in our sample have observations at the $K$ and $Q$ bands, reliable results are obtained only for the $S$ and $X$ bands.
At the $S$ band, the median structure index is about 2, while at the $X$ band it is about 3. 
The median compactness of the sample is about 0.6 in both the $S$ and $X$ bands.
We find that the structure index of our sample shows no significant correlation with redshift in any band. 

\begin{figure*}
  \centering
  \includegraphics[width=0.9\textwidth]{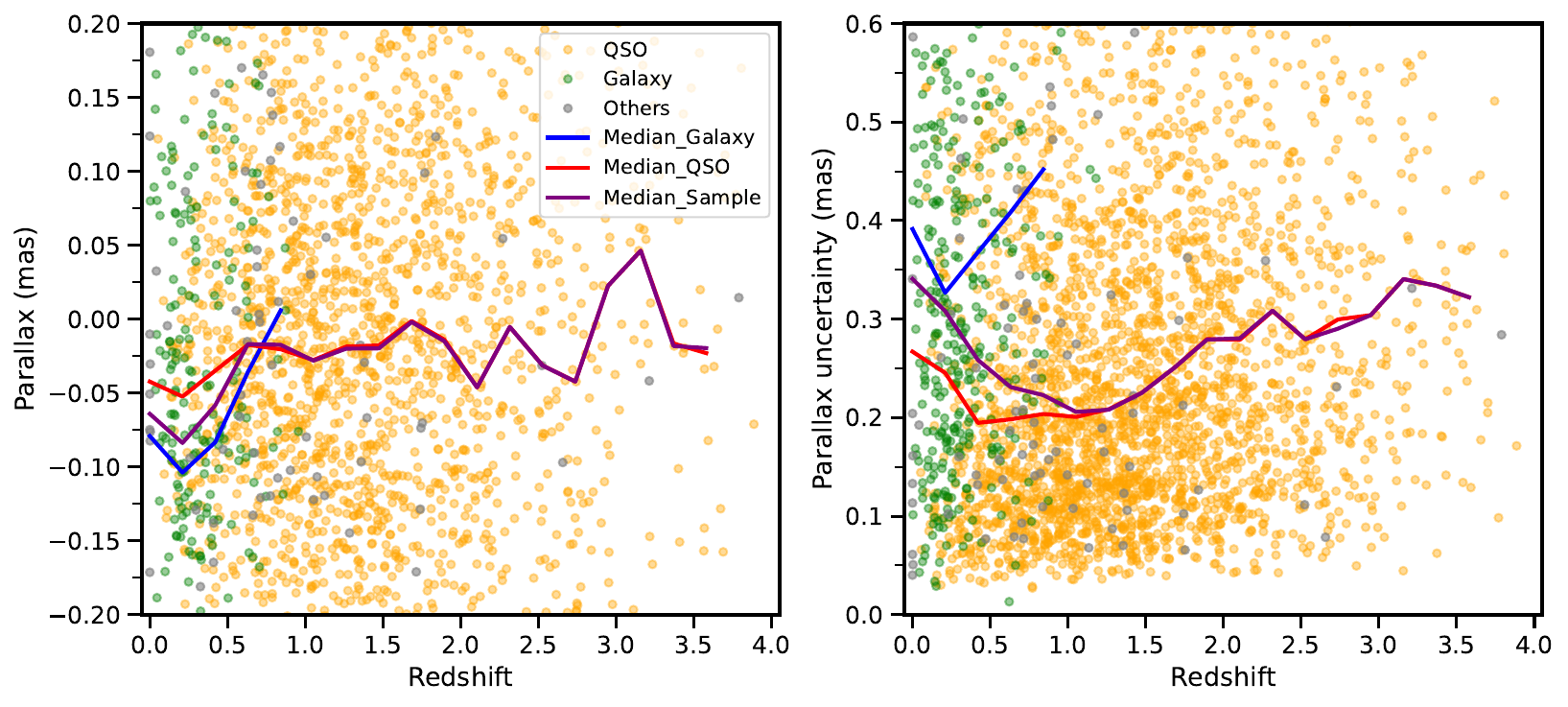}  
  \caption[]{\label{fig:PLX_uncertainty} 
  Parallax (left) and parallax uncertainty (right) as functions of redshift.
  The meanings of the colours, data points, and lines are the same as those in Fig.~\ref{fig:uncer}.
  }
\end{figure*}

\begin{figure*}
  \centering
  \includegraphics[width=0.9\textwidth]{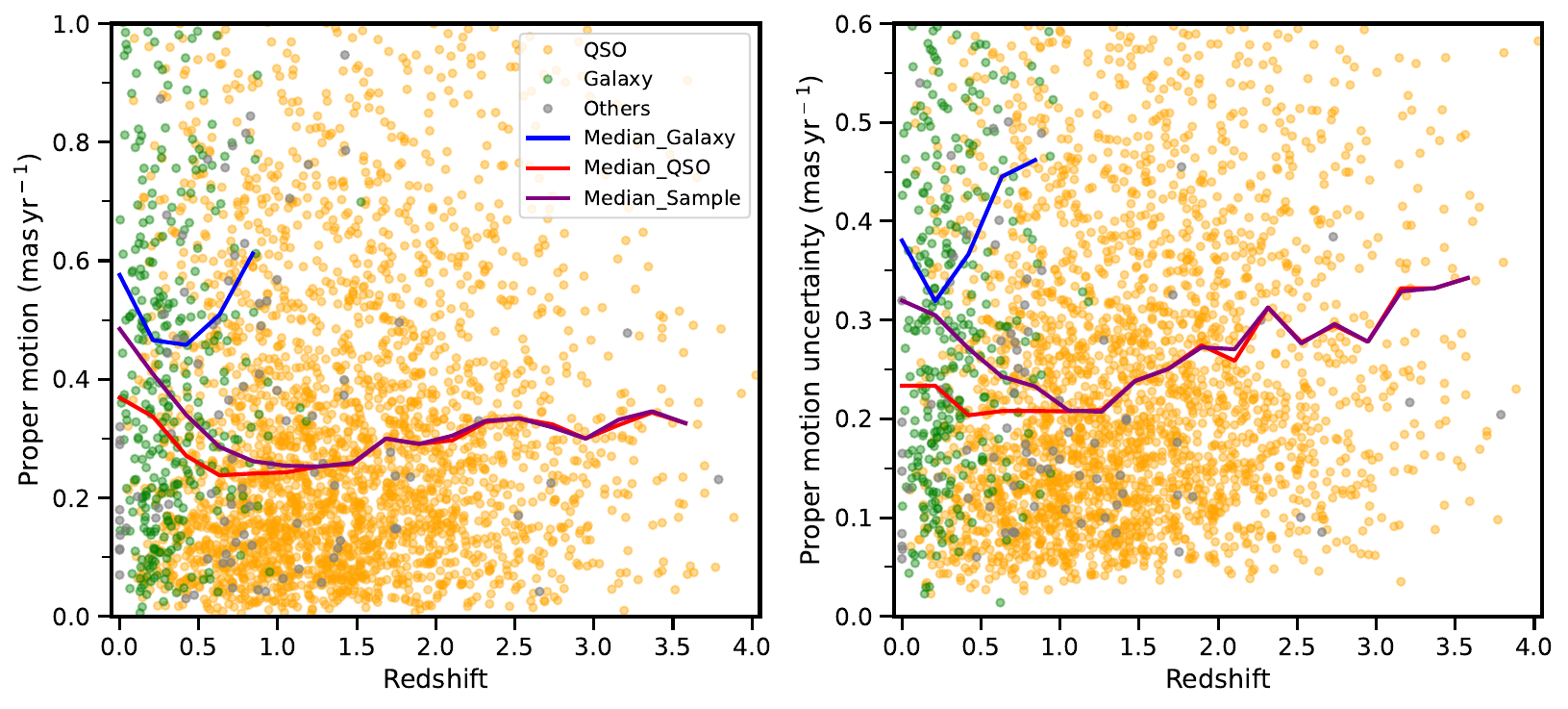}  
  \caption[]{\label{fig:PM_uncertainty} 
  Proper motion (left) and proper motion uncertainty (right) as functions of redshift.
  The meanings of the colours, data points, and lines are the same as those in Fig.~\ref{fig:uncer}.
  }
\end{figure*}

\subsection{Radio-optical offset}    \label{sec:Offset}
We calculated absolute and relative offsets following the same procedures as in \citetads{2016A&A...595A...5M}, where these offsets are referred to as the angular and normalised separations, respectively.
As shown in Fig.~\ref{fig:offsets}, the median absolute and relative radio-optical offsets of the total sample decrease rapidly with increasing redshift over the range $0\!<z\!<0.5$, followed by a slower decrease over $0.5\!<z\!<1.3$.
The median absolute offset reaches a minimum around $z\!\simeq\!1.3$ and exhibits a slight increase over the range $1.3\!<z\!<3.0$ (from 0.7~mas at $z\!\simeq\!1.3$ to 1.2~mas at $z\!\simeq\!3.0$).
However, the median relative offset remains nearly constant beyond $z\!>\!1.3$.
This overall trend of absolute and relative offsets is consistent with the results reported by \citetads{2019ApJ...873..132M} for ICRF3 sources.
Additionally, this behaviour is similar to the trends observed in the optical positional uncertainties (Fig.~\ref{fig:uncer}).

This similarity suggests that the observed trend largely reflects the changing composition of the sample.
Galaxies dominate the total sample at low redshift ($0\!<z\!<0.5$). 
They not only exhibit systematically larger absolute offsets than QSOs, but also show a strong decrease in their offsets with increasing redshift.
As a result, the median offsets of the total sample decrease rapidly over $0\!<z\!<0.5$.
Over the range $0.5\!<z\!<1$, the fraction of galaxies becomes smaller and the offsets of the total sample decrease more slowly.
At higher redshift ($z\!>\!1$), the total sample is dominated by QSOs, whose absolute offsets show only a weak increase with redshift, while the relative offsets remain roughly constant.
The combination of these population fractions and their distinct redshift behaviours results in the overall trend observed for the total sample.

\begin{figure*}
  \centering
  \includegraphics[width=0.9\textwidth]{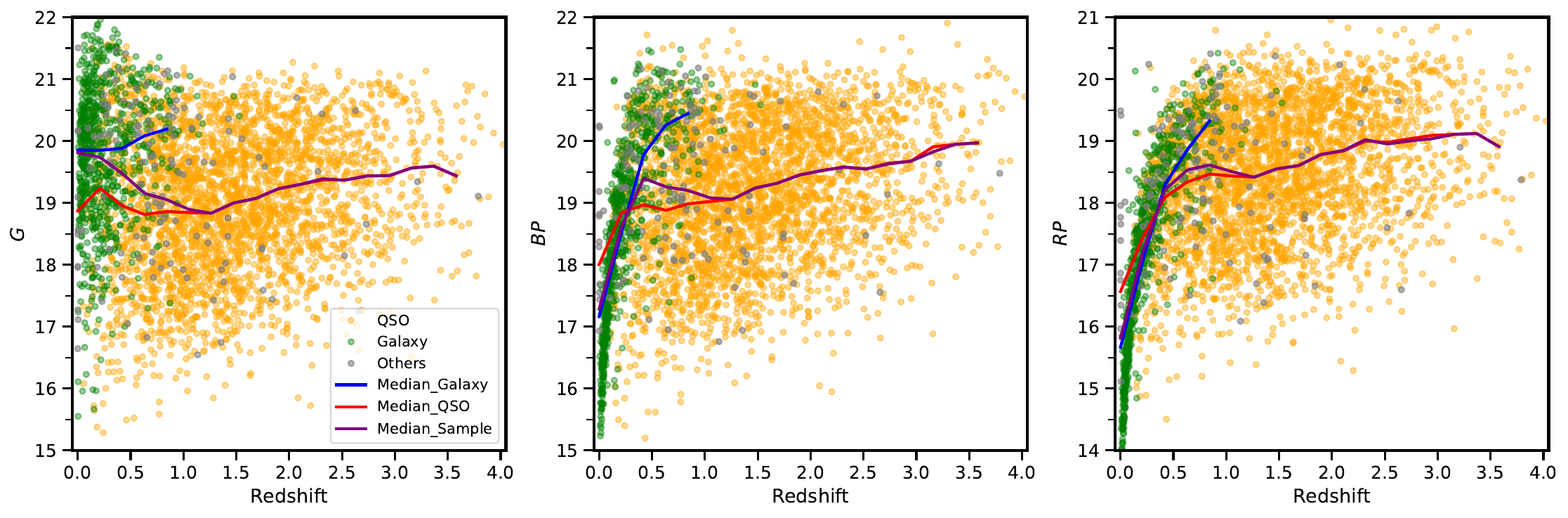}  
  \caption[]{\label{fig:mag} 
  $G$ (left), $BP$ (middle), and $RP$ (right) magnitudes as functions of redshift.
  The meanings of the colours, data points, and lines are the same as those in Fig.~\ref{fig:uncer}.
  }
\end{figure*}

\begin{figure*}
  \centering
  \includegraphics[width=0.9\textwidth]{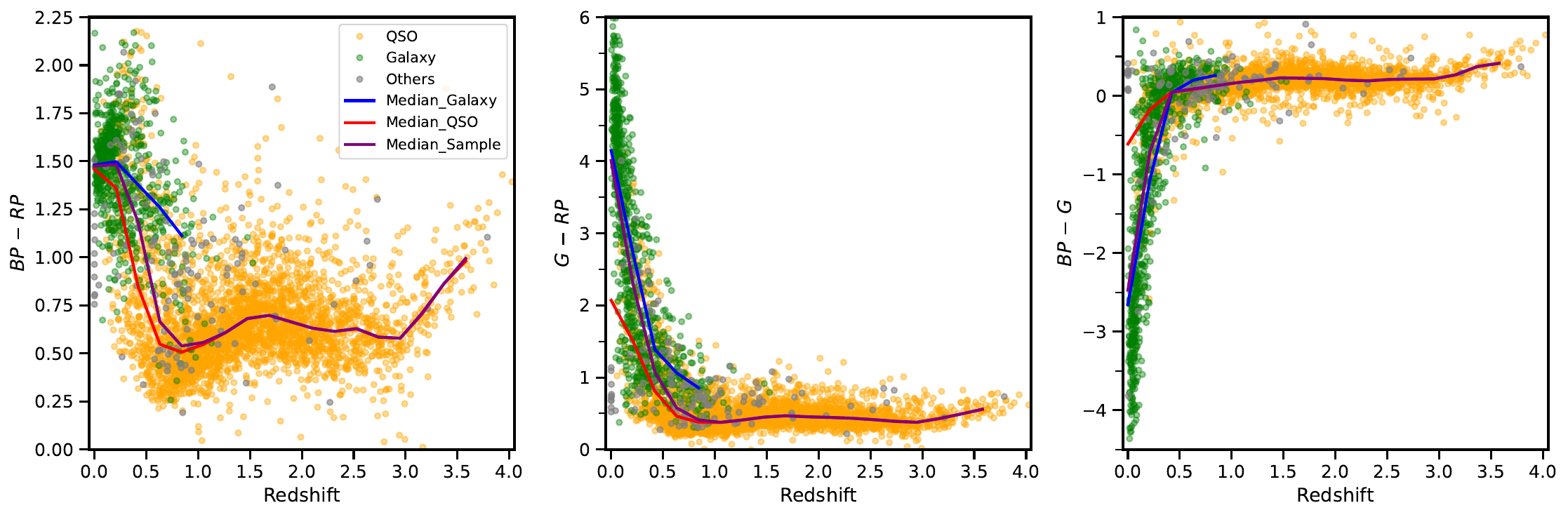}  
  \caption[]{\label{fig:CI} 
  $BP - RP$ (left), $G - RP$ (middle), and $BP - G$ (right) colour indices as functions of redshift.
  The meanings of the colours, data points, and lines are the same as those in Fig.~\ref{fig:uncer}.
  }
\end{figure*}

\subsection{Parallaxes and proper motions}    \label{sec:Optical}

As shown in Fig.~\ref{fig:PLX_uncertainty}, we investigated the relationships between parallax and redshift. 
We find that the parallax of the sample shows no clear dependence on redshift, with the median values in different bins concentrated between 0 and $-0.1$~mas.
For QSOs, the median parallax remains around $-0.02$~mas over the redshift range $0.5\!<z\!<1.5$, which likely reflects the parallax zero-point offset present in \textit{Gaia} measurements (\citeads{2021A&A...649A...4L}; \citeads{2024A&A...691A..81D}). 
In both the total sample and the QSO subsample, the parallax uncertainty decreases significantly with increasing redshift over the range $0\!<z\!<0.5$, followed by a slower decrease over $0.5\!<z\!<1$. For $z\!>\!1$, the parallax uncertainty increases slowly with redshift, rising from about $0.2$~mas at $z\!\sim\!1$ to about $0.3$~mas at $z\!\sim\!3.5$.
In addition, compared with QSOs at comparable redshifts, galaxies exhibit a more negative median parallax and larger parallax uncertainties.

We also investigated the relationships between proper motion and redshift (Fig.~\ref{fig:PM_uncertainty}). 
We find a similar behaviour to that of the optical positional uncertainties: both the median proper motion and its median uncertainty decrease rapidly over $0\!<z\!<0.5$, followed by a slower decrease over $0.5\!<z\!<1$, and exhibit only a slight increase at $z\!>\!1$. 
This trend is consistent with the changing source composition, in which galaxies dominate at low redshift and exhibit larger values that decrease clearly with increasing redshift, whereas QSOs dominate at high redshift ($z\!>\!1$) and exhibit smaller values with only a weak increase.

Furthermore, we computed the error-normalised proper motion $\chi$ as in \citetads{2022ApJ...933...28M}:
\begin{equation}
\chi^2\!=\!\frac{1}{1-\rho^2} \left[\left(\frac{\mu_{\alpha*}}{\sigma_{\mu_{\alpha*}}}\right)^2 + \left(\frac{\mu_{\delta}}{\sigma_{\mu_{\delta}}}\right)^2 - 2\rho \frac{\mu_{\alpha*}\mu_{\delta}} {\sigma_{\mu_{\alpha*}}\sigma_{\mu_{\delta}}} \right],
\end{equation}
where $\mu_{\alpha*}$ and $\mu_{\delta}$ are the \textit{Gaia} proper motions in right ascension and declination, with uncertainties $\sigma_{\mu_{\alpha*}}$ and $\sigma_{\mu_{\delta}}$, and correlation coefficient $\rho$.
The median error-normalised proper motion of the total sample decreases from 1.6 at $z\!\simeq\!0$ to 1.2 at $z\!\simeq\!0.5$, and remains roughly constant at 1.2 at redshifts $z\!>\!0.5$.
This behaviour is consistent with the expectation for a Rayleigh distribution with $\sigma\!=\!1$, whose median is 1.18.
Within the statistical precision of our relatively small RFC-based sample, the intrinsic proper motions of most sources with $z\!>\!0.5$ are consistent with zero.
Our result is therefore in agreement with the results obtained from a much larger \textit{Gaia}~EDR3-based sample \citepads{2022ApJ...933...28M}, which indicate that the observed \textit{Gaia} proper motions are on average dominated by random measurement uncertainties.

\subsection{Photometric properties}    \label{sec:Photometric}

The main differences between different types of RFC sources, based on their spectral classification, are most clearly seen in the relationship between magnitude and redshift.
As shown in Fig.~\ref{fig:mag}, we examined the relationships between redshift and the three \textit{Gaia} magnitudes ($G$, $BP$, and $RP$) of the sample. 
We find that the median $G$ magnitude of galaxies remains roughly constant at $\sim\!20$ and shows no significant dependence on redshift, while the median $G$ magnitude of QSOs is around $\sim\!19$ at the same redshift. 
Therefore, although the $G$ magnitudes of both galaxies and QSOs show no strong redshift dependence individually, the concentration of galaxies within $0\!<z\!<1$ combined with their systematically fainter $G$ magnitudes compared to QSOs results in a noticeable negative gradient in the median $G$ magnitude of the total sample over this redshift range.
In contrast, the $BP$ and $RP$ magnitudes of galaxies exhibit a strong dependence on redshift. 
Within $0\!<z\!<1$, the median $BP$ magnitude of galaxies increases sharply from $\sim\!17$ at $z\!=\!0$ to $\sim\!20.5$ at $z\!=\!1$, while the median $RP$ magnitude increases from $\sim\!16$ to $\sim\!19.5$ over the same redshift range.
However, QSOs exhibit only a slight increase with increasing redshift in $G$, $BP$ and $RP$ magnitudes.
This difference is reflected in the median $BP$ and $RP$ magnitudes of the total sample. 
They increase rapidly over $0\!<z\!<0.5$, followed by a slower increase over $0.5\!<z\!<1$. 
However, at $z\!>\!1$, the sample is dominated by QSOs and the median magnitudes show only a weak increase with redshift.

The relationships between $G$, $BP$, and $RP$ magnitudes and redshift also influence the evolution of the corresponding colour indices with redshift. 
As shown in Fig.~\ref{fig:CI}, we examined the relationships between redshift and three colour indices: $BP - RP$ (left), $G - RP$ (middle), and $BP - G$ (right). 
These colour indices exhibit relatively tighter distributions compared to the magnitudes.
We find that the median $BP - RP$ colour of the total sample decreases sharply with increasing redshift over $0\!<z\!<1$, remains nearly constant at $\sim\!0.6$ over $1\!<z\!<3$ (with a slight increase to $\sim\!0.7$ near $z\!\simeq\!1.5$), and rises significantly at $z\!>\!3$. 
This trend is consistent with that of the QSOs.
For galaxies, the median $BP - RP$ remains around $\sim\!1.5$ within $0\!<z\!<0.5$, which likely contributes to the negative gradient observed in the total sample over this redshift range.
The redshift dependence of the median $G - RP$ and $BP - G$ colours is mainly determined by the strong redshift dependence of the median $BP$ and $RP$ magnitudes in galaxies, whereas the median $G$ magnitude remains nearly constant over $0\!<z\!<1$.
As a result, the median $G - RP$ colour of galaxies decreases sharply and the median $BP - G$ colour increases sharply with redshift within this range.
For $z\!>\!1$, the median $G - RP$ colour of QSOs is concentrated around $\sim\!0.4$ while the median $BP - G$ colour is concentrated around zero, and neither of them shows significant dependence on redshift. 
Since galaxies dominate the sample at low redshift ($0\!<z\!<0.5$), the total sample exhibits strong gradients in the median $G - RP$ and $BP - G$ colours over this range, consistent with the trends of galaxies. 
Over $0.5\!<z\!<1$, the fraction of galaxies decreases and the evolution becomes more gradual. 
At higher redshift ($z\!>\!1$), the sample is dominated by QSOs and the median values remain nearly constant, at $\sim\!0.4$ and $\sim\!0.1$, respectively.
It is worth noting that the median $G - RP$ colour of QSOs shows a slight enhancement near $z\!\simeq\!1.5$ and a modest increase at $z\!>\!3$, consistent with the trend of the median $BP - RP$ colour with redshift.

\section{Discussion} \label{sec:Discussion}
\subsection{Comparison with existing redshift catalogues}
In the RFC cross-match table, redshift information is available for approximately 8\,900 sources, most of which were compiled from NED. 
These redshifts originate from a mixture of spectroscopic and photometric measurements, as well as machine-learning estimates.
About 2\,500 sources (28\%) are based on SDSS~DR13 spectroscopy \citepads{2017ApJS..233...25A}, while an additional 713 sources (8\%) rely on photometric redshifts from SDSS~DR6 \citepads{2009ApJS..180...67R} and the Two Micron All Sky Survey \citepads[2MASS;][]{2006AJ....131.1163S}, which are subject to relatively large uncertainties.
In this work, we substantially expanded and updated the spectroscopic redshift content by incorporating the latest measurements from SDSS~DR17 and DR19, and in particular from DESI~DR1 \citepads[released in March 2025;][]{2025arXiv250314745D}. 
As a result, we obtain a sample of 4\,774 RFC sources with reliable spectroscopic redshifts, accounting for more than 50\% of the redshifts in the RFC cross-match table.
During the compilation process, we carefully considered redshift quality indicators (e.g. {\tt ZWARN} and {\tt ZWARNING}) and performed cross-checks between independent measurements from SDSS and DESI. 
This procedure ensures a reliable spectroscopic redshift sample, forming the basis of our analysis of redshift-dependent astrometric properties.

In similar previous studies 
(\citeads{2019ApJ...873..132M}; \citeads{2021A&A...652A..87L}; \citeads{2025MNRAS.542.2389H}), redshift information was primarily drawn from catalogues such as OCARS and LQAC. 
In addition to ground-based SDSS~DR16 spectroscopy, the most recent OCARS \citepads{2025AJ....170...48M} and LQAC-6 \citepads{2024A&A...683A.112S} releases also include redshifts from the \textit{Gaia}~DR3 Quasar Source Classifier \citepads[QSOC;][]{2023A&A...674A..31D}.
However, \citetads{2025MNRAS.542.2389H} reported significant offsets relative to ground-based spectroscopy for 37.8\% of QSOC redshifts.
At the time of submission of this work, neither NED, OCARS, nor LQAC-6 included redshifts from the newly released DESI~DR1 or SDSS~DR19, although NED did contain redshifts from the earlier DESI EDR \citepads{2024AJ....168...58D}.

The RFC cross-match table also provides source classifications primarily based on NED.
We refined these classifications using spectroscopic information from SDSS and DESI, adopting a simplified scheme that separates RFC sources into galaxies and QSOs, while excluding objects with uncertain or missing classifications. 
This approach reduces classification complexity and facilitates a clearer comparison of redshift-dependent astrometric behaviour between the two populations.
In future work, we plan to implement a more detailed spectroscopic classification, similar to that of \citetads{2022ApJS..260...33S}, by identifying subclasses such as Seyfert~I, Seyfert~II, and blazars, and to include this information in an updated RFC redshift catalogue.

\subsection{Comparison with previous studies}

The median absolute and relative radio–optical offsets decrease significantly with increasing redshift over the range $0\!<z\!<1$, while the absolute offset reaches a minimum at $z\!\simeq\!1.3$ (Fig.~\ref{fig:offsets}). 
This overall decrease at low redshifts is consistent with the findings of \citetads{2019ApJ...873..132M}, who reported a similar decline up to $z\!\simeq\!1.5$ based on 2\,074 ICRF3 radio-loud quasars cross-matched with \textit{Gaia}~DR2 positions and OCARS redshifts. 
The modest difference between the turnover redshifts identified in our study ($z\!\simeq\!1.3$) and in theirs ($z\!\simeq\!1.5$) is well within the expected uncertainties and likely reflects differences in sample selection and binning strategies.
At higher redshifts, \citetads{2019ApJ...873..132M} found that the relative radio–optical offset becomes approximately constant, while the absolute offset increases steadily for $z\!>\!1.6$.
This result is consistent with our findings, in that the absolute offset exhibits a slight increase over the range $1.3\!<z\!<3.0$, while the relative offset remains nearly constant beyond $z\!>\!1$ in our sample.
They attributed the rise in absolute offsets at high redshift to a combination of instrumental and astrophysical effects, including the systematic fading of $G$ magnitudes with increasing redshift and the entrance of the C\,IV emission line into the \textit{Gaia} $G$ band at $z\!>\!1.6$, which may shift the optical photocentre relative to the radio core. 
In our analysis, we find a similar redshift dependence of the $G$ magnitudes (Fig.~\ref{fig:mag}). 
At $z\!<\!1.3$, the presence of host galaxies causes the $G$ magnitudes of the total sample to become significantly brighter with increasing redshift, whereas at $z\!>\!1.3$, both the QSOs and the full sample exhibit a mild fading trend, in agreement with \citetads{2025A&A...695A.135L}.
In addition, we find that the C\,IV emission line fully enters the \textit{Gaia} $G$ band (wavelength range 330--1050~nm) by $z\!\sim\!1.48$, as illustrated in Fig.~\ref{fig:Spectra}(b). 
This implies that the C\,IV line begins to contribute appreciably to the $G$-band flux already at $z\!\simeq\!1.3$, slightly earlier than the $z\!\simeq\!1.6$ threshold suggested by \citetads{2019ApJ...873..132M}.
This earlier onset may partly explain the slightly lower turnover redshift found in our analysis.

Additionally, the redshift dependence of the $BP - RP$ colour in our sample (Fig.~\ref{fig:CI}) is consistent with the trend reported by \citetads{2019ApJ...873..132M}.
Over $0\!<z\!<1$, the $BP - RP$ colour of both QSOs and the total sample decreases significantly with increasing redshift. This behaviour reflects the diminishing contribution of red galaxies at higher redshift, together with the fact that the intrinsically stronger UV emission of QSOs is progressively redshifted into the $BP$ band.
At intermediate redshifts, a slight increase in $BP - RP$ around $z\!\simeq\!1.5$ may be related to the Mg\,II emission line entering the $RP$ band.
As shown in Fig.~\ref{fig:Spectra}, Mg\,II appears at $\sim\!690$\,nm at $z\!\sim\!1.48$, just beyond the upper wavelength limit of the $BP$ band (330--680\,nm) and within the $RP$ band (640--1050\,nm), consistent with the interpretation of \citetads{2019ApJ...873..132M}.
At high redshift ($z\!>\!3$), the $BP - RP$ colour increases sharply for Ly$\alpha$-forest QSOs. 
In this regime, the $BP$ band is strongly affected by Ly$\alpha$-forest absorption: as redshift increases, the Ly$\alpha$ break shifts redward across the $BP$ band, causing a growing fraction of the band to be absorbed and the $BP$ magnitude to become significantly fainter than the $RP$ magnitude. 
This effect naturally explains the pronounced reddening observed at $z\!>\!3$.

Similarly, \citetads{2021A&A...652A..87L} investigated the optical-to-radio offsets at the $S/X$, $K$, and $X/Ka$ bands as a function of redshift for a sample of 456 ICRF3 sources with redshift measurements from the LQAC-5 catalogue.
They found no evidence for a dependence of the optical-to-radio offset on the redshift.
There are 934 ICRF3 sources in our sample, and the radio-optical offset as a function of redshift obtained from this subset exhibits a trend that is consistent with that of the total sample as shown in Fig.~\ref{fig:offsets}.
The difference between our results and theirs may be explained by the much smaller number of sources in their sample, which limits the statistical robustness of their result.
They also reported a clear dependence of both absolute and relative optical–to–radio offsets on the \textit{Gaia} $G$ magnitude: absolute offsets increase toward fainter sources, while relative offsets decrease with increasing $G$ magnitude at all three radio bands. 
The increase of absolute offsets toward fainter $G$ magnitudes is consistent with our findings. 
However, unlike \citetads{2021A&A...652A..87L}, we do not find evidence for a significant dependence of the relative radio-optical offsets on $G$ magnitude (Fig.~\ref{fig:offsets_G}).

Using \textit{Gaia}~DR3 astrometry for quasars in LQAC-6, \citetads{2025MNRAS.542.2389H} showed that both the optical positional and proper-motion uncertainties follow an exponential dependence on $G$ magnitude, consistent with the \textit{Gaia} astrometric performance reported by \citetads{2023A&A...674A...1G}. 
In agreement with these results, we find that the \textit{Gaia} positional uncertainties (Fig.~\ref{fig:uncer_G}), parallax uncertainties (Fig.~\ref{fig:PLX_G}), and proper-motion uncertainties (Fig.~\ref{fig:PM_G}) in our sample all increase steeply toward fainter $G$ magnitudes and are well described by exponential relations.
Moreover, the measured proper motions exhibit a $G$-magnitude dependence that closely mirrors that of their uncertainties. 
This similarity indicates that the observed proper motions of RFC sources are, on average, dominated by random measurement uncertainties rather than intrinsic source motions.

Overall, we find that the \textit{Gaia} optical positional uncertainties, parallax uncertainties, proper motions and their uncertainties, $G$ magnitudes, and both absolute and relative optical–radio offsets of RFC sources exhibit a consistent redshift dependence: they decrease rapidly over $0\!<z\!<0.5$, decline more gradually over $0.5\!<z\!<1$, and show little variation or only a weak increase at $z\!>\!1$.
At the same time, most of these quantities increase exponentially toward fainter $G$ magnitudes.
This strongly indicates that the apparent redshift dependence of these quantities mainly reflects their fundamental dependence on $G$ magnitude, which is the dominant factor governing \textit{Gaia}'s astrometric precision.

We therefore conclude that the redshift evolution of \textit{Gaia} astrometric uncertainties is primarily a consequence of the redshift-dependent behaviour of the $G$ magnitudes of RFC sources.
The approximately constant median error-normalised proper motion ($\sim$1.2) at $z\!>\!0.5$ further indicates that the observed \textit{Gaia} proper motions in this regime are dominated by random measurement uncertainties. 
Similarly, the redshift dependence of the absolute optical–radio offsets is mainly driven by the behaviour of the optical positional uncertainties.

Physically, the evolution of $G$-band magnitude with redshift can arise from several factors, including changes in the dominant source population (e.g., a transition from galaxies to QSOs), the redshifting of strong spectral features into or out of the $G$ band (such as C\,IV or Ly$\alpha$), and distance-related selection effects. 
These factors naturally produce systematic variations in the $G$ magnitude with redshift, which in turn govern the observed redshift dependence of \textit{Gaia} astrometric parameters.

\subsection{Difference between QSOs and galaxies}
In this work, we separate RFC sources into galaxies and QSOs to investigate differences in the redshift evolution of their astrometric properties.
At the same redshift, galaxies are systematically fainter than QSOs, with median $G$ magnitudes of $\sim\!20$ and $\sim\!19$, respectively (Fig.~\ref{fig:mag}). 
Since the $G$ magnitude is the dominant factor determining \textit{Gaia} astrometric precision \citepads{2023A&A...674A...1G}, this systematic faintness leads galaxies to exhibit larger optical astrometric uncertainties and consequently larger radio-optical positional offsets.
In addition, galaxies show more negative measured parallaxes and larger proper motions than QSOs at the same redshift, reflecting increased measurement noise rather than intrinsic motions.

The photometric evolution further highlights the distinction between these two populations.
For galaxies, both $BP$ and $RP$ magnitudes increase rapidly with redshift (Fig.~\ref{fig:mag}) because these two bands correspond to bluer rest-frame wavelengths where galaxies are much fainter and this effect becomes stronger when the $4000~\AA$ break shifts into the $BP$ or $RP$ bands.
Thus, the flux in these bands drops sharply, causing the corresponding magnitudes to rise rapidly.
In contrast, the spectra of QSOs are well described by a power law and remain bright in the UV band, resulting in much weaker redshift dependence in these bands. 
This effect is less pronounced in the broad $G$ band because its wide wavelength coverage smooths over these spectral variations.

In addition to photometric effects, source morphology may also play an important role.
As extended systems, galaxies are susceptible to photocentre shifts induced by structural features such as spiral arms, dust lanes, bulges, bars, tidal tails, or interaction-driven asymmetries, which can significantly increase astrometric uncertainties and optical–radio offsets \citepads{2026ApJ...997..283K}.
QSOs intrinsically exhibit morphological structures such as jets, accretion disks, and in some cases, double or multiple systems that can induce photocenter shifts (\citeads{2020ApJ...888...73H}; \citeads{2021A&A...651A..64L}).
However, most QSO sources are typically unresolved and effectively point-like in \textit{Gaia} observations, making their measured centroids less sensitive to morphological structures than those of extended galaxies and rendering them more reliable astrometric reference sources \citepads{2022A&A...667A.148G}.
This difference is directly reflected in Fig.~\ref{fig:uncer_G} where galaxies show larger optical positional uncertainties than QSOs at the same $G$ magnitude, further resulting in larger parallax and proper motion uncertainties, as well as radio–optical offsets. 

Finally, classification effects may partially blur the distinction between the two populations at low redshift.
Some galaxies exhibit unusual broad emission-line components that may be misclassified as QSOs by automated spectral pipelines.
Such misclassifications can make the statistical properties of QSOs at low redshift ($0\!<z\!<0.5$) appear more similar to those of galaxies. 
In addition, genuine AGNs (including QSOs) at low redshift ($0\!<z\!<0.5$) often have partially resolved host galaxies, which introduce additional extended emission that complicates centroid determination and increases astrometric uncertainties.
A more robust classification of galaxies and QSOs will therefore require future work involving careful spectra inspection and improved spectral fitting techniques.
 
\section{Conclusion} \label{sec:Conclusion}
We have combined DESI~DR1 and SDSS~DR17/DR19 spectroscopy with \textit{Gaia}~DR3 to construct a large and homogeneous sample of RFC sources with reliable redshifts and classifications. 
Using this sample, we investigate the redshift dependence of astrometric parameters from VLBI observations and \textit{Gaia}~DR3, providing additional observational support for the consistency between the optical and radio celestial reference frames.

The radio-related quantities, including radio positional uncertainties, structure index, and compactness, show no significant dependence on redshift within the current precision. 
In contrast, \textit{Gaia} astrometric parameters exhibit clear redshift-dependent behaviour. 
For the total sample, optical positional uncertainties, proper-motion uncertainties, radio–optical offsets, and the $G$ magnitude decrease rapidly over $0\!<z\!<0.5$, decline more slowly over $0.5\!<z\!<1$, and remain approximately constant at higher redshift, reflecting the transition from galaxy-dominated to QSO-dominated populations.

The absolute radio–optical offsets show an exponential dependence on $G$ magnitude and are systematically larger for galaxies than for QSOs. 
While radio positional uncertainties depend only weakly on $G$ magnitude, optical positional uncertainties increase exponentially toward fainter sources, indicating that radio–optical offsets are primarily driven by optical astrometric errors. 
Parallaxes show no significant dependence on either redshift or $G$ magnitude; for QSOs, the median parallax remains close to $-0.02$~mas, consistent with the known \textit{Gaia} parallax zero-point offset. 
Proper motions and their uncertainties exhibit an exponential dependence on $G$ magnitude, and the nearly constant median error-normalised proper motion ($\sim\!1.2$) at $z\!>\!0.5$ indicates that the observed proper motions are dominated by random measurement uncertainties.

Galaxies are systematically fainter than QSOs, with median $G$ magnitudes of $\sim\!20$ and $\sim\!19$, respectively, and therefore show larger \textit{Gaia} astrometric uncertainties at a given redshift. 
Even at fixed $G$ magnitude, galaxies exhibit larger uncertainties, implying an additional contribution from their extended morphology. 
Their $BP$ and $RP$ magnitudes fade rapidly with redshift as the 4000~$\AA$ break shifts into these bands, whereas QSOs show much weaker photometric evolution. 
Consequently, the colour evolution of the sample is governed by population changes at low redshift and by spectral-feature shifts at higher redshift. 
The $BP - RP$ colour decreases over $0\!<z\!<1$, shows a mild increase near $z\!\simeq\!1.5$ due to Mg\,II entering the $RP$ band, and rises sharply at $z\!>\!3$ when the Ly$\alpha$ break crosses the $BP$ band, extending previous studies \citepads{2019ApJ...873..132M} to higher redshift.

In future, we will construct a more comprehensive redshift catalogue for RFC sources by using additional spectroscopic surveys and performing a more detailed classification.
We will further complement these data with photometric redshifts and machine-learning estimates where spectroscopic measurements are unavailable.
This will enable a more robust characterisation of redshift-dependent astrometric effects and further strengthen the optical–radio reference frame connection.

\begin{acknowledgements}
  We sincerely thank the anonymous referee, S\'ebastien Lambert, Leonid Petrov, Jun Yang, Xiaopeng Cheng and Bo Zhang for the fruitful discussions and useful suggestions, which have greatly improved the work.
  
  This work was supported by the National Natural Science Foundation of China (NSFC) under grant Nos.~12573070 and 12373074.

  We acknowledge the use of the Radio Fundamental Catalogue \citepads{2025ApJS..276...38P}.
  This work has made use of data from the European Space Agency (ESA) mission \textit{Gaia} (\url{https://www.cosmos.esa.int/gaia}), processed by the \textit{Gaia} Data Processing and Analysis Consortium (DPAC, \url{https://www.cosmos.esa.int/web/gaia/dpac/consortium}). 
  Funding for the DPAC has been provided by national institutions, in particular the institutions participating in the \textit{Gaia} Multilateral Agreement. 

  Funding for the Sloan Digital Sky Survey V has been provided by the Alfred P. Sloan Foundation, the Heising-Simons Foundation, the National Science Foundation, and the Participating Institutions. 
  SDSS acknowledges support and resources from the Center for High-Performance Computing at the University of Utah.
  SDSS telescopes are located at Apache Point Observatory, funded by the Astrophysical Research Consortium and operated by New Mexico State University, and at Las Campanas Observatory, operated by the Carnegie Institution for Science. 
  The SDSS web site is www.sdss.org.

  This research used data obtained with the Dark Energy Spectroscopic Instrument (DESI).
  DESI construction and operations are managed by the Lawrence Berkeley National Laboratory. 
  This material is based upon work supported by the U.S. Department of Energy, Office of Science, Office of High-Energy Physics, under Contract No. DE–AC02–05CH11231, and by the National Energy Research Scientific Computing Center, a DOE Office of Science User Facility under the same contract. 
  Additional support for DESI was provided by the U.S. National Science Foundation (NSF), Division of Astronomical Sciences under Contract No. AST-0950945 to the NSF’s National Optical-Infrared Astronomy Research Laboratory; the Science and Technology Facilities Council of the United Kingdom; the Gordon and Betty Moore Foundation; the Heising-Simons Foundation; the French Alternative Energies and Atomic Energy Commission (CEA); the National Council of Humanities, Science and Technology of Mexico (CONAHCYT); the Ministry of Science and Innovation of Spain (MICINN), and by the DESI Member Institutions: \url{www.desi.lbl.gov/collaborating-institutions}.
  The DESI collaboration is honored to be permitted to conduct scientific research on I’oligam Du’ag (Kitt Peak), a mountain with particular significance to the Tohono O’odham Nation. 
  Any opinions, findings, and conclusions or recommendations expressed in this material are those of the author(s) and do not necessarily reflect the views of the U.S. National Science Foundation, the U.S. Department of Energy, or any of the listed funding agencies.

  This research has made use of the NASA/IPAC Extragalactic Database (NED), which is funded by the National Aeronautics and Space Administration and operated by the California Institute of Technology.

  This research used material from the Bordeaux VLBI Image Database (BVID). 
  This database can be reached at \url{http://bvid.astrophy.u-bordeaux.fr/}.
  
  We made much use of Astropy (\url{http://www.astropy.org}) – a community-developed core Python package for astronomy \citepads{2018AJ....156..123A}, the Python 2D plotting library Matplotlib \citepads{2007CSE.....9...90H} and TOPCAT \citepads{2011ascl.soft01010T}.
  
  This work benefited from the LaTeX service provided by the e-Science Center of the Collaborative Innovation Center of Advanced Microstructures, Nanjing University (\url{https://sci.nju.edu.cn}).

  Researchers interested in the sample-related data may contact us via email.
\end{acknowledgements}
\bibliographystyle{aa}
\bibliography{aa}
\newpage
\begin{appendix}
\onecolumn
\section{Examples of DESI spectra}\label{appendix:desi-spectra}

\begin{figure}[ht!]
  \centering
  \includegraphics[width=0.88\textwidth]{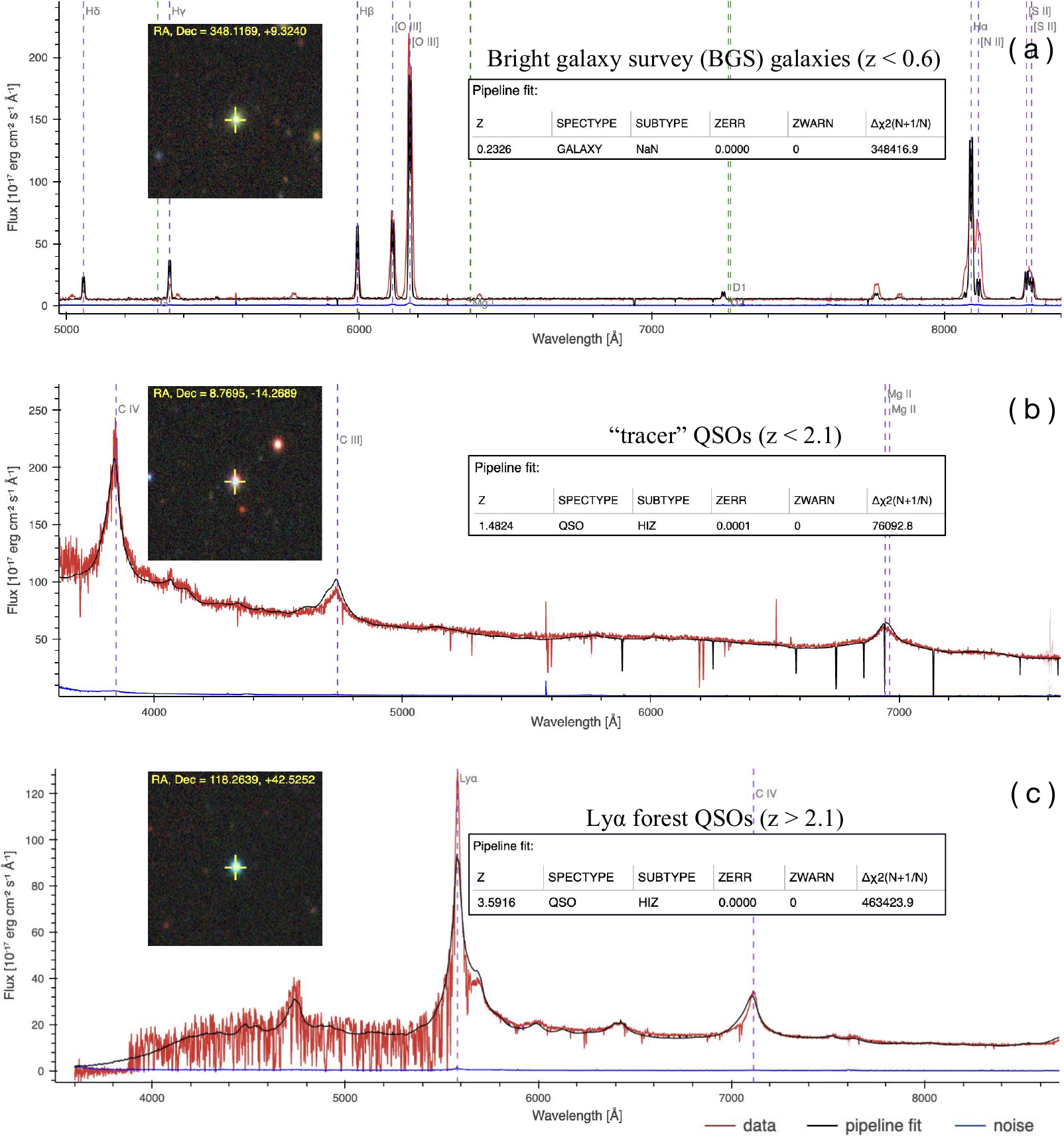}  
  \caption[]{\label{fig:Spectra} 
  Spectra of three example sources. Panel a: A Bright galaxy survey (BGS) galaxy exhibiting Balmer emission lines ($z\!<\!0.6$). Panel b: A ``tracer'' QSO showing Mg\,II, C\,III] and C\,IV lines ($z\!<\!2.1$). Panel c: A Ly$\alpha$ forest QSO with a prominent Ly$\alpha$ emission line and associated Ly$\alpha$ forest absorption ($z\!>\!2.1$).  
  }
\end{figure}

DESI targets five primary classes of objects, including Milky Way Survey (MWS) and backup-program stars, bright galaxies from the Bright Galaxy Survey (BGS; $0\!<z\!<0.6$), luminous red galaxies (LRGs; $0.4\!<z\!<1.1$), emission-line galaxies (ELGs; $0.6\!<z\!<1.6$), and quasars (QSOs; $0.9\!<z\!<4$). 
For cosmological analyses, DESI further divides the QSO sample into “tracer” QSOs at $z\!<\!2.1$ and Ly$\alpha$ forest QSOs at $z\!>\!2.1$, with the latter used as an independent probe of the matter-density field through absorption features in the Ly$\alpha$ forest.
We show DESI spectra and their high-precision redshift measurements for three example sources in Fig.~\ref{fig:Spectra}: a BGS galaxy exhibiting Balmer emission lines (panel a), a “tracer” QSO showing Mg\,II, C\,III] and C\,IV lines (panel b), and a QSO showing Ly$\alpha$ emission line with Ly$\alpha$ forest (panel c).
In each panel, a DESI Legacy Surveys DR9 image cutout of the source is shown on the left, together with the observed spectrum (red), the DESI pipeline best-fit spectrum (black), and the noise estimate (blue), while the source information derived from the pipeline fits is summarised in a table on the right.
The table lists the redshift and its uncertainty, the redshift quality flag (${\tt ZWARN}$), the primary and secondary spectral classifications, and the $\Delta\chi^2$ statistic from the pipeline fitting.

\section{Cross-checking of redshifts from different methods} \label{appendix:z-crosscheck}

\begin{figure}[ht!]
  \centering
  \includegraphics[width=0.8\textwidth]{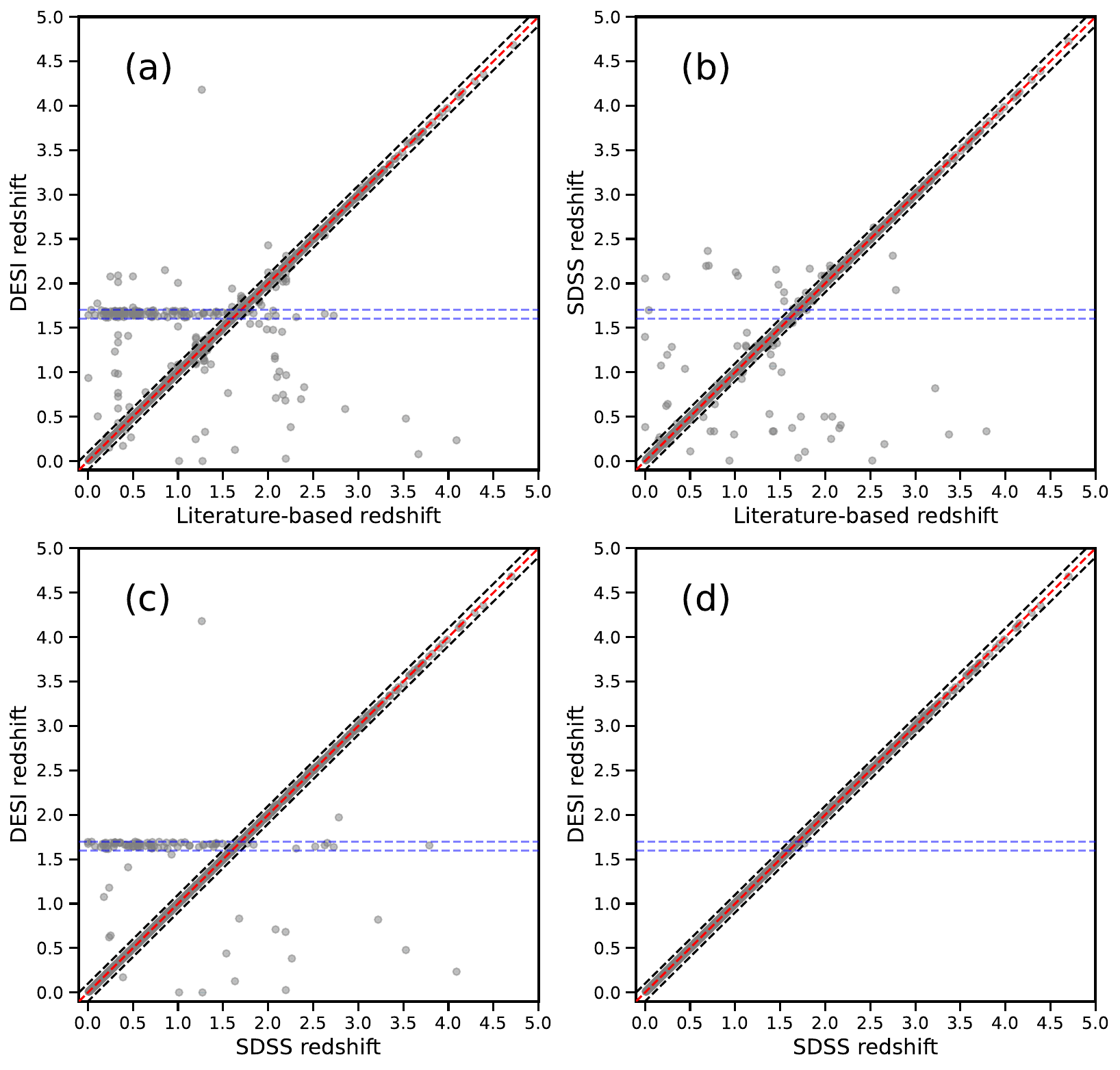}  
  \caption[]{\label{fig:crosschecking} 
  Cross-checking of redshifts derived from different methods. 
  Panel a: Spectroscopic redshifts from DESI versus redshifts from other literature. 
  Panel b: Spectroscopic redshifts from SDSS versus redshifts from other literature. 
  Panel c: Spectroscopic redshifts from DESI versus spectroscopic redshifts from SDSS. 
  Panel d: Spectroscopic redshifts from DESI versus spectroscopic redshifts from SDSS after excluding data points with offsets larger than 0.1, representing the most reliable subset of our final sample. 
  In each panel, grey points represent the original data, the red diagonal dashed line shows the 1:1 relation.  
  The two blue dashed lines mark the region $1.6\!<z\!<1.7$ on the vertical axis, where DESI redshift measurements show known systematic issues. 
  The two black dashed lines mark the region where $|\Delta z|\!<\!0.1$; most data points fall within this region, indicating that the redshifts from the two independent measurements are largely consistent.
  }
\end{figure}

As shown in Fig.~\ref{fig:crosschecking}, we performed cross-checking of redshifts between different methods. 
DESI spectroscopic redshifts are compared with the literature-based redshifts (panel a), and SDSS spectroscopic redshifts from DR17 and DR19 (panel c). 
Similarly, SDSS spectroscopic redshifts are cross-checked against the literature-based redshifts (panel b). 
We find that the majority of sources show good agreement in redshifts across different methods ($|\Delta z|\!<\!0.1$).
However, we find that DESI spectroscopic redshifts in the range $1.6\!<z\!<1.7$ show significant deviations from redshifts measured by other methods (panels a and c).
In this work, for sources with both SDSS and DESI spectroscopic redshifts, we retained only those with consistent redshifts satisfying $|\Delta z|\!<\!0.1$ (panel d).

\section{Astrometric properties as a function of $G$ magnitude} \label{appendix:G magnitude}
\begin{figure*}
  \centering
  \includegraphics[width=0.9\textwidth]{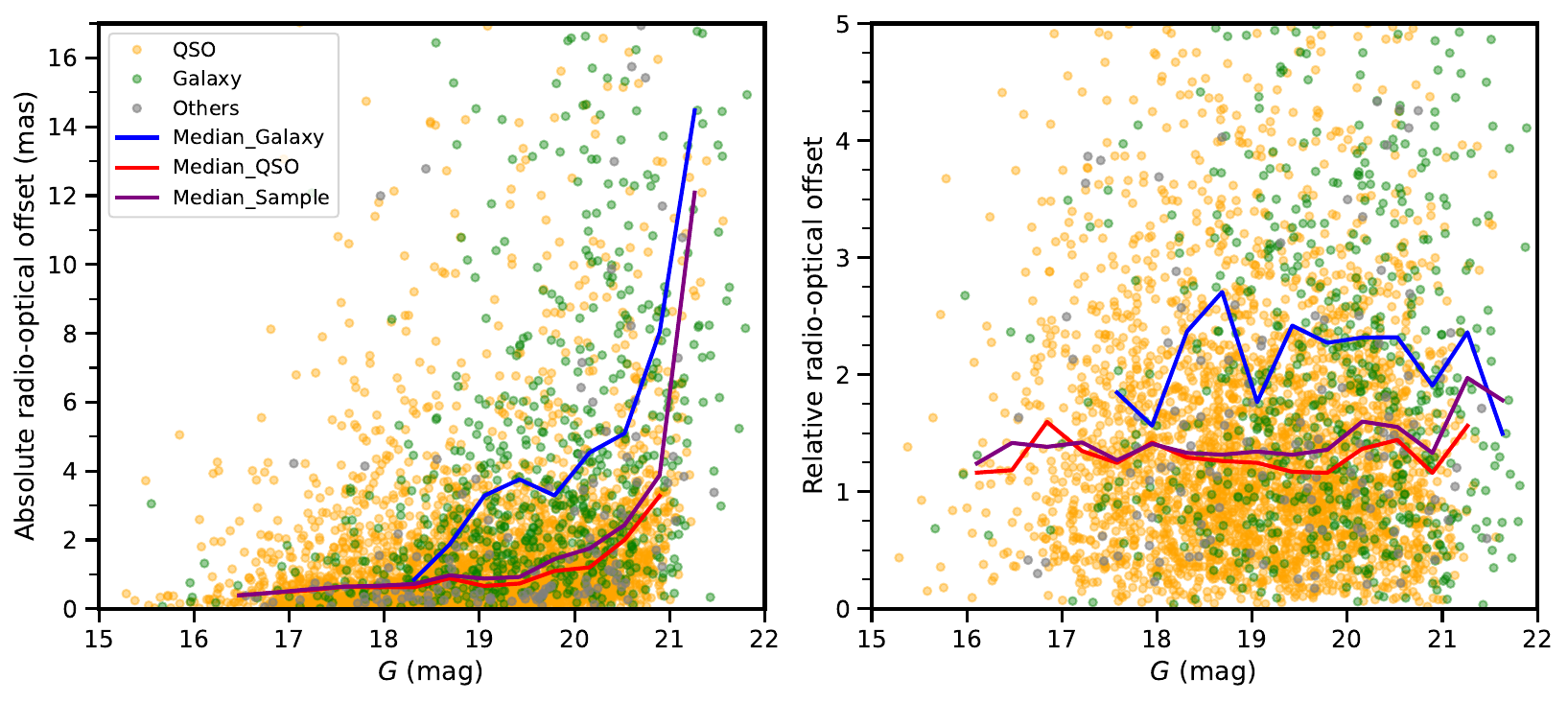}  
  \caption[]{\label{fig:offsets_G} 
  Absolute radio-optical offset (left) and relative radio-optical offset (right) as functions of $G$ magnitude.
  The meanings of the colours, data points, and lines are the same as those in Fig.~\ref{fig:uncer}.
  Median values are not shown for bins containing fewer than 10 data points.
  }
\end{figure*}

\begin{figure*}
  \centering
  \includegraphics[width=0.9\textwidth]{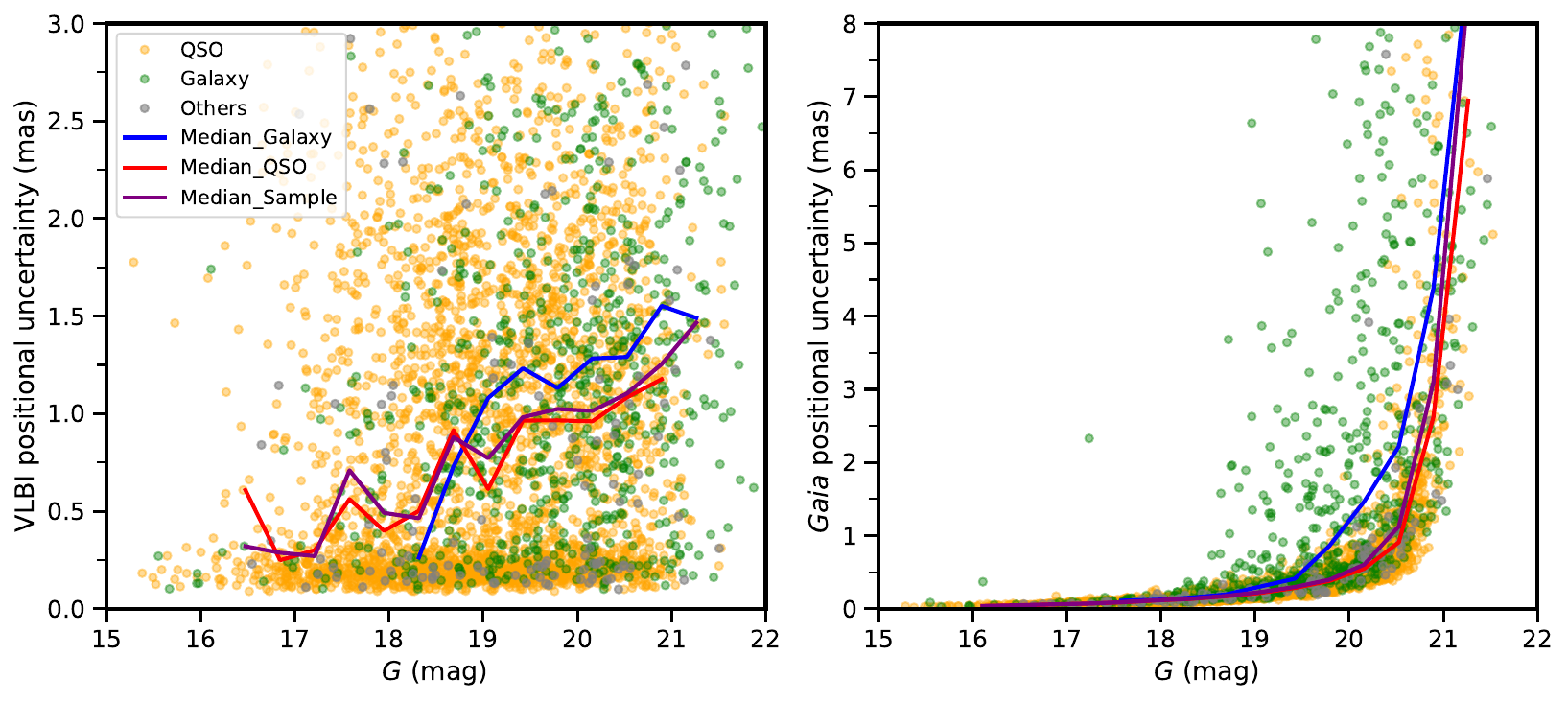}  
  \caption[]{\label{fig:uncer_G} 
  Positional uncertainty derived from RFC VLBI observations (left) and \textit{Gaia} optical measurements (right) as functions of $G$ magnitude.
  The meanings of the colours, data points, and lines are the same as those in Fig.~\ref{fig:uncer}. 
  Median values are not shown for bins containing fewer than 10 data points.
  }
\end{figure*}

\begin{figure*}
  \centering
  \includegraphics[width=0.9\textwidth]{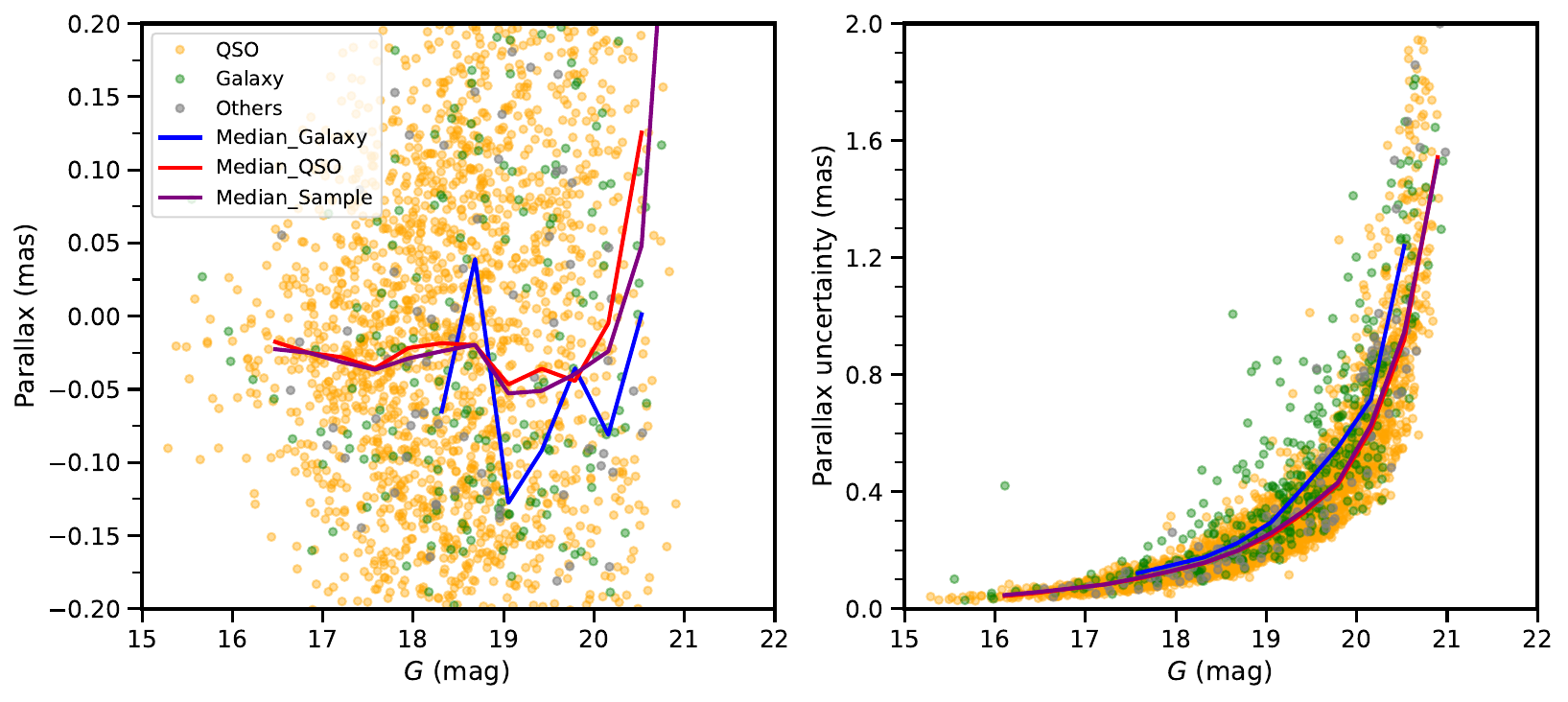}  
  \caption[]{\label{fig:PLX_G} 
  Parallax (left) and parallax uncertainty (right) as functions of $G$ magnitude.
  The meanings of the colours, data points, and lines are the same as those in Fig.~\ref{fig:uncer}. 
  Median values are not shown for bins containing fewer than 10 data points.
  }
\end{figure*}

\begin{figure*}
  \centering
  \includegraphics[width=0.9\textwidth]{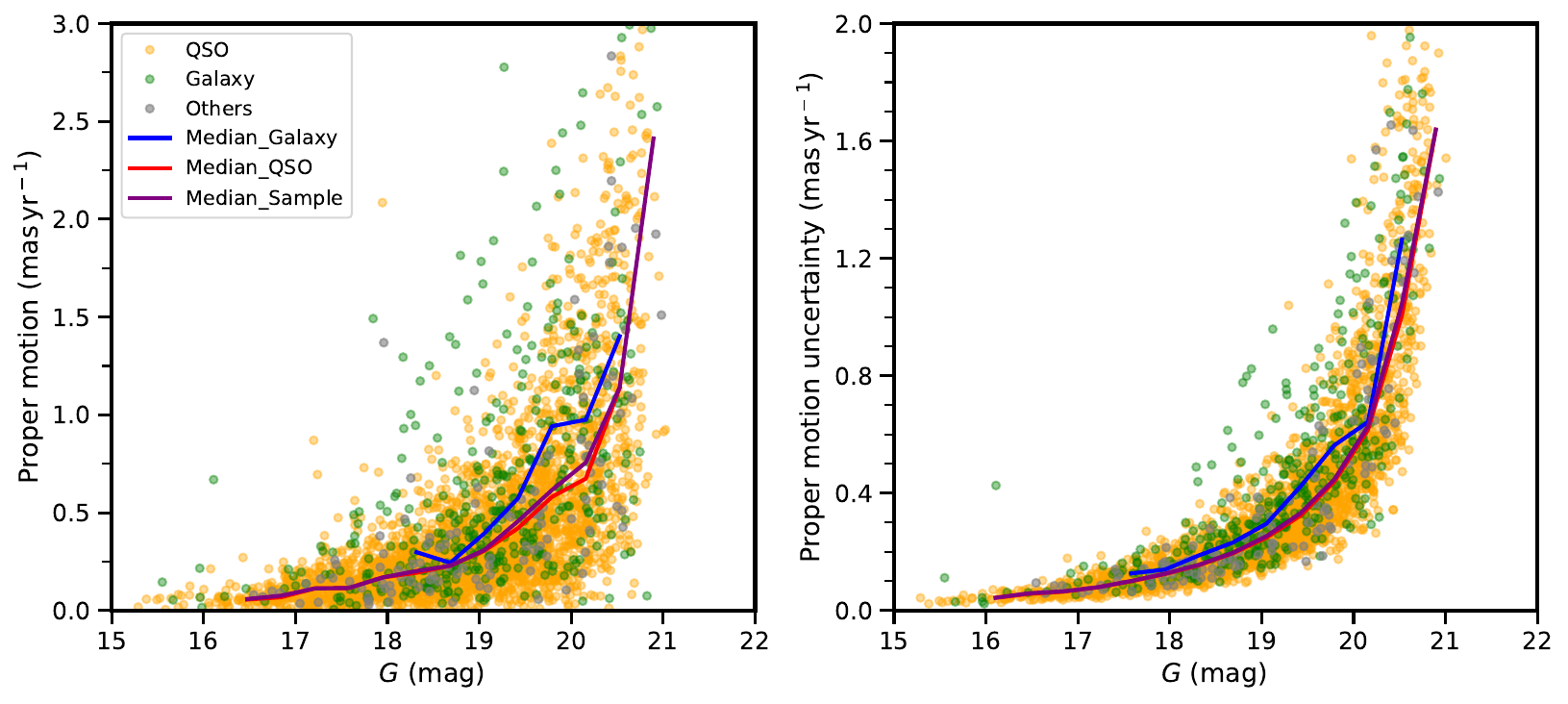}  
  \caption[]{\label{fig:PM_G} 
  Proper motion (left) and proper motion uncertainty (right) as functions of $G$ magnitude.
  The meanings of the colours, data points, and lines are the same as those in Fig.~\ref{fig:uncer}. 
  Median values are not shown for bins containing fewer than 10 data points.
  }
\end{figure*}

In this work, we find that the absolute radio-optical offsets (Fig.~\ref{fig:offsets_G}), the \textit{Gaia} optical positional uncertainties (Fig.~\ref{fig:uncer_G}), parallax uncertainties (Fig.~\ref{fig:PLX_G}), proper motions and proper motion uncertainties (Fig.~\ref{fig:PM_G}) in our sample all increase steeply toward fainter $G$ magnitudes and can be well described by an exponential dependence on $G$.
In contrast, as shown in Fig.~\ref{fig:uncer_G}, the median radio positional uncertainties increase only mildly and approximately linearly with $G$ magnitude, from about 0.5~mas at $G\!\sim\!18$ to roughly 1.5~mas at $G\!\sim\!21$, which is significantly smaller than the corresponding optical positional uncertainties at faint $G$ magnitudes. 
However, at a given $G$ magnitude, both the absolute and relative radio–optical offsets are significantly larger for galaxies than for QSOs. 
The \textit{Gaia} optical positional uncertainties are also significantly larger for galaxies than for QSOs, while the radio positional uncertainties are comparable between galaxies and QSOs.
Taken together, these results indicate that the observed radio–optical offsets of RFC sources are most likely dominated by uncertainties in the \textit{Gaia} optical positions rather than in the VLBI radio measurements.

\end{appendix}
\end{document}